\RequirePackage{ifpdf}
\ifpdf 
\documentclass[pdftex]{sigma}
\else
\documentclass{sigma}
\fi

\begin{document}

\numberwithin{equation}{section}

\renewcommand{\PaperNumber}{076}

\FirstPageHeading

\ShortArticleName{SU$_2$ Nonstandard Bases: Case of Mutually
Unbiased Bases}

\ArticleName{SU$_2$ Nonstandard Bases:\\ Case of Mutually Unbiased
Bases}

\Author{Olivier ALBOUY and Maurice R. KIBLER}

\AuthorNameForHeading{O. Albouy and M.R. Kibler}

\Address{Universit\'e de Lyon, Institut de Physique Nucl\'eaire,
Universit\'e Lyon 1 and CNRS/IN2P3,\\ 43 bd du 11 novembre 1918,
F--69622 Villeurbanne Cedex, France}
\Email{\href{mailto:o.albouy@ipnl.in2p3.fr}{o.albouy@ipnl.in2p3.fr},
\href{mailto:m.kibler@ipnl.in2p3.fr}{m.kibler@ipnl.in2p3.fr}}

\ArticleDates{Received April 07, 2007, in f\/inal form June 16,
2007; Published online July 08, 2007}

\Abstract{This paper deals with bases in a f\/inite-dimensional
Hilbert space. Such a~space can be realized as a subspace of the
representation space of SU$_2$ corresponding to an irreducible
representation of SU$_2$. The representation theory of SU$_2$ is
reconsidered via the use of two truncated deformed oscillators.
This leads to replacement of the familiar scheme $\{ j^2 , j_z \}$
by a scheme $\{ j^2 , v_{ra} \}$, where the two-parameter operator
$v_{ra}$ is def\/ined in the universal enveloping algebra of the
Lie algebra su$_2$. The eigenvectors of the commuting set of
operators $\{ j^2 , v_{ra} \}$ are adapted to a tower of chains
SO$_3 \supset C_{2j+1}$ ($2j \in \mathbb{N}^{\ast}$), where
$C_{2j+1}$ is the cyclic group of order $2j+1$. In the case where
$2j+1$ is prime, the corresponding eigenvectors generate a
complete set of mutually unbiased bases. Some useful relations on
generalized quadratic Gauss sums are exposed in three appendices.}


\Classification{81R50; 81R05; 81R10; 81R15}

\section{Introduction}

Utilisation of linear combinations of simultaneous eigenstates
$\left\vert jm\right\rangle$ of the square $j^{2}$ and the
component $j_{z}$ of a generalized angular momentum is widespread
in physics. For instance, in molecular physics and condensed
matter physics, we employ state vectors of the type
          \begin{gather}
\left\vert ja\Gamma \gamma \right\rangle
=\sum_{m=-j}^{j}\left\vert jm\right\rangle \left\langle
jm|ja\Gamma \gamma \right\rangle, \label{Decomposition jm}
          \end{gather}
where $\Gamma$ stands for an irreducible representation of a
subgroup $G^{\ast}$ of the group SU$_2$, $\gamma$ is a~label for
dif\/ferentiating the various partners of $\Gamma$ (in the case
where $\dim \Gamma \geq 2$) and $a$ denotes a~multiplicity label
necessary when $\Gamma$ occurs several times in the irreducible
representation~$(j)$ of~SU$_2$. The group $G^{\ast}$ is the spinor
or double group of a (generally f\/inite) subgroup $G$ of SO$_3$,
$G$~being a point group of molecular or crystallographic interest.
The vectors (\ref{Decomposition jm}) are referred to as symmetry
adapted vectors or symmetry adapted functions in the context of
molecular orbital theory \cite{1Melvin}.

The state vector $\left\vert ja\Gamma \gamma \right\rangle$ is an
eigenvector of $j^{2}$ and of the projection operator
          \begin{gather}
P_{\gamma }^{\Gamma }=\frac{\dim \Gamma }{\left\vert G^{\ast
}\right\vert } \sum_{R\in G^{\ast }}\overline{D^{\Gamma
}(R)_{\gamma \gamma }}P_{R}, \label{Projection operators}
          \end{gather}
where the bar indicates complex conjugation and $\left\vert
G^{\ast }\right\vert$ is the order of $G^{\ast}$. Here, we use
$D^{\Gamma }(R)_{\gamma \gamma ^{\prime }}$ to denote the $\gamma
\gamma ^{\prime}$ matrix element of the matrix representation
$D^{\Gamma}$ associated with $\Gamma $. In addition, the operator
$P_{R}$ acts on $\left\vert ja\Gamma \gamma \right\rangle $ as
          \begin{gather}
P_{R}\left\vert ja\Gamma \gamma \right\rangle =\sum_{\gamma
^{\prime }=1}^{\dim \Gamma }D^{\Gamma }(R)_{\gamma ^{\prime
}\gamma }\left\vert ja\Gamma \gamma ^{\prime }\right\rangle.
\label{Action of PR}
          \end{gather}
Note that it is always possible to assume that the matrix
representation $D^{\Gamma}$ in (\ref{Projection operators}) and
(\ref{Action of PR}) is unitary. Thus for f\/ixed $j$, the vectors
$\left\vert j a\Gamma \gamma \right\rangle$ are eigenvectors of
the set
  $\{ j^2, P_{\gamma }^{\Gamma }   :   \gamma = 1, 2, \dots, \dim \Gamma \}$
of commuting operators.

In the situation where $j$ is an integer, realizations of
(\ref{Decomposition jm}) on the sphere $S^{2}$ are known as
$G$-harmonics and play an important role in chemical physics and
quantum chemistry \cite{2Altmann-1, 2Altmann-2}. More generally,
state vectors (\ref{Decomposition jm}) with $j$ integer or half of
an odd integer are of considerable interest in electronic
spectroscopy of paramagnetic ions in f\/inite symmetry
\cite{3Kibler-1, 3Kibler-2} and/or in rotational-vibrational
spectroscopy of molecules \cite{4Moret-1, 4Moret-2}.

It is to be noted that in many cases the labels $a$ and $\gamma $
can be characterized (at least partially) by irreducible
representations of a chain of groups containing $G^{\ast}$ and
having SU$_2$ as head group. In such cases, the state vectors
(\ref{Decomposition jm}) transform according to irreducible
representations of the groups of the chain under consideration.

It is also possible to give more physical signif\/icance to the
label $a$. For this purpose, let us consider an operator $v$
def\/ined in the enveloping algebra of SU$_2$ and invariant under
the group $G$. The operators $v$ and $j^{2}$ obviously commute.
According to Wigner's theorem, the eigenvectors of $j^{2}$ and $v$
are of type (\ref{Decomposition jm}), where $a$ stands for an
eigenvalue of $v$. In that case, the label $a$ may be replaced by
an eigenvalue $\lambda$ of $v$. (We assume that there is no state
labeling problem, i.e., the triplets $\lambda \Gamma \gamma $
completely label the state vectors within the irreducible
representation $(j)$ of SU$_2$.) We are thus led to supersymmetry
adapted vectors of the type
          \[
\left\vert j\lambda \Gamma \gamma \right\rangle
=\sum_{a}\left\vert ja\Gamma \gamma \right\rangle U_{a\lambda },
          \]
where the unitary matrix $U$ diagonalizes the matrix $v$ set up on
the set $\{\left\vert ja\Gamma \gamma \right\rangle : a \ {\rm
ranging}\}$. Integrity bases to obtain $v$ were given for
dif\/ferent subgroups $G$ of SO$_3$ \cite{5Patera-1, 5Patera-2,
5Patera-3}.

As a r\'{e}sum\'{e}, there are several kinds of physically
interesting bases for the irreducible representations of SU$_2$.
The standard basis, associated with the commuting set
$\{j^2,j_z\}$, corresponds to the canonical group-subgroup chain
SU$_2 \supset {\rm U}_1$. For this chain, we have $\Gamma := m$
and there is no need for the labels $a$ and $\gamma $. Another
kind of group-subgroup basis can be obtained by replacing U$_1$ by
another subgroup $G^{\ast}$ of the group SU$_2$. Among the various
SU$_2 \supset G^{\ast}$ symmetry adapted bases (SABs), we may
distinguish (i) the weakly SABs $\{\left\vert ja\Gamma \gamma
\right\rangle :a,\Gamma ,\gamma \ {\rm ranging}\}$ for which the
symmetry adapted vectors are eigenvectors of $j^2$ and of the
projection operators of $G^{\ast}$ and (ii) the strongly SABs
$\{\left\vert ja\Gamma \gamma \right\rangle :\lambda ,\Gamma
,\gamma \ {\rm ranging}\}$ for which the supersymmetry adapted
vectors are eigenvectors of $j^2$ and of an operator def\/ined in
the enveloping algebra of SU$_2$ and invariant under the group
$G$. Both for strongly and weakly SABs, the restriction of SU$_2$
to $G^{\ast}$ yields a decomposition of the irreducible
representation $(j)$ of SU$_2$ into a direct sum of irreducible
representations $\Gamma $ of $G^{\ast}$.

It is the object of the present work to study nonstandard bases of
SU$_2$\ with a tower of chains SU$_2 \supset C_{2j+1}^{\ast}$ or
                       SO$_3 \supset C_{2j+1}$, with $2j\in
\mathbb{N}^{\ast}$, where $C_{2j+1}$ is the cyclic group of order
$2j+1$. In other words, the chain of groups used here depends on
the irreducible representation of SU$_2$ to be considered. We
shall establish a connection, mentioned in \cite{6KibPla}, between
the obtained bases and the so-called mutually unbiased bases
(MUBs) used in quantum information.

The organisation of this paper is as follows. In Section 2, we
construct the Lie algebra of SU$_2$ from two quon algebras $A_{1}$
and $A_{2}$\ corresponding to the same deformation parameter $q$
taken as a root of unity. Section 3 deals with an alternative to
the $\{j^2,j_z\}$ scheme, viz., the $\{j^2,v_{ra}\}$ scheme, which
corresponds, for f\/ixed $j$, to a set of polar decompositions of
SU$_2$ with $a=0, 1, \dots, 2j$. Realizations of the operators
$v_{ra}$ in the enveloping algebra of SU$_2$ are given in
Section~3. The link with MUBs is developed in Section~4. In a
series of appendices, we give some useful relations satisf\/ied by
generalized quadratic Gauss sums.

\section[A quon realization of the algebra su$_2$]{A quon realization of the algebra su$\boldsymbol{{}_2}$}

\subsection{Two quon algebras}

Following the works in \cite{7quons-1, 7quons-2, 7quons-3,
7quons-4, 7quons-5}, we def\/ine two quon algebras $A_i = \{
a_{i-}, a_{i+}, N_i \}$ with $i = 1$ and $2$ by
          \begin{gather*}
   a_{i-}a_{i+} - qa_{i+}a_{i-} = 1,            \qquad
   \left[ N_i, a_{i\pm} \right] = \pm a_{i\pm}, \qquad
   N_i^{\dagger} = N_i,                         \qquad
  \left( a_{i\pm} \right)^k = 0,
\\
  \forall \; x_1 \in A_1, \ \forall \; x_2 \in A_2 : \ [x_1 , x_2] = 0,
\end{gather*}
where
          \begin{gather}
     q = \exp \left( {2 \pi {i} \over k} \right), \quad
     k \in \mathbb{N} \setminus \{ 0,1 \}.
    \label{definition of q}
          \end{gather}
The generators $a_{i \pm}$ and $N_i$ of $A_i$ are linear
operators. As in the classical case $q=1$, we say that $a_{i +}$
is a creation operator, $a_{i -}$ an annihilation operator and
$N_i$ a number operator. Note that the case $k=2$ corresponds to
fermion operators and the case $k \to \infty$ to boson operators.
In other words, each of the algebras $A_i$ describes fermions for
$q=-1$ and bosons for $q=1$. The nilpotency conditions $\left(
a_{i\pm} \right)^k = 0$ can be understood as describing a
generalized exclusion principle for particles of fractional spin
$1/k$ (the Pauli exclusion principle corresponds to $k = 2$). Let
us mention that algebras similar to $A_1$ and $A_2$ with $N_1 =
N_2$ were introduced in \cite{8DaoHasKib-1, 8DaoHasKib-2,
8DaoHasKib-3} to def\/ine $k$-fermions which are, like anyons,
objects interpolating between fermions (corresponding to $k = 2$)
and bosons (corresponding to $k \to \infty$).

\subsection{Representation of the quon algebras}

We can f\/ind several Hilbertian representations of the algebras
$A_1$ and $A_2$. Let ${\cal F}(i)$ be two truncated Fock--Hilbert
spaces of dimension $k$ corresponding to two truncated harmonic
oscillators ($i=1,2$). We endow each space ${\cal F}(i)$ with an
orthonormalized basis $\{ | n_i ) : n_i = 0, 1, \dots, k-1 \}$. As
a generalization of the representation given in
\cite{8DaoHasKib-1, 8DaoHasKib-2, 8DaoHasKib-3}, we have the
following result.

\begin{proposition}
The relations
          \begin{gather}
  a_{1+} |n_1) = \left( \left[ n_1 + s + \frac{1}{2} \right]_q \right)^a |n_1 + 1), \qquad
  a_{1+} |k-1) = 0,
  \label{action de a1+}
\\
  a_{1-} |n_1) = \left( \left[ n_1 + s - \frac{1}{2} \right]_q \right)^c |n_1 - 1), \qquad
  a_{1-} |0)   = 0,
  \label{action de a1-}
\\
  a_{2+} |n_2) = \left( \left[ n_2 + s + \frac{1}{2} \right]_q \right)^b |n_2 + 1), \qquad
  a_{2+} |k-1) = 0,
  \label{action de a2+}
\\
  a_{2-} |n_2) = \left( \left[ n_2 + s - \frac{1}{2} \right]_q \right)^d |n_2 - 1), \qquad
  a_{2-} |0) = 0,
  \label{action de a2-}
          \end{gather}
and
          \[
  N_1 |n_1) = n_1 |n_1),                          \qquad
  N_2 |n_2) = n_2 |n_2)
          \]
define a family of representations of $A_1$ and $A_2$ depending on
two independent parameters, say~$a$ and~$b$, with $a + c = b + d =
1$.
\end{proposition}

In equations~(\ref{action de a1+})--(\ref{action de a2-}), we take
$s=1/2$. Furthermore, we use
          \[
  \left[ x \right]_q = \frac{1-q^x}{1-q}, \qquad x \in \mathbb{R}.
          \]
We shall also use the $q$-deformed factorial def\/ined by
          \[
  \left[ n \right]_q! =
  \left[ 1 \right]_q
  \left[ 2 \right]_q  \cdots
  \left[ n \right]_q, \qquad n \in \mathbb{N}^{\ast},
                                     \qquad \left[ 0 \right]_q! = 1.
          \]

We continue with the representation of $A_1 \otimes A_2$
af\/forded by $a=0$ and $b=1$. The operators on $A_1 \otimes A_2$
act on the f\/inite-dimensional Hilbert space ${\cal F}_k = {\cal
F}(1) \otimes {\cal F}(2)$ of dimension $k^2$. The set $\{  | n_1
, n_2 ) = | n_1) \otimes | n_2 ) : n_1 , n_2 = 0, 1, \dots, k-1
\}$ constitute an orthonormalized basis of ${\cal F}_k$. We denote
$( \ | \ )$ the scalar product on ${\cal F}_k$ so that
          \[
( n_1' , n_2' | n_1 , n_2 ) = \delta_{n_1' , n_1 } \>
                              \delta_{n_2' , n_2 }.
          \]

\subsection{Two basic operators}
Following \cite{9KibDao-1, 9KibDao-2}, we def\/ine the two linear
operators
          \begin{gather}
  h = {\sqrt {N_1 \left( N_2 + 1 \right) }}, \qquad v_{ra} = s_1 s_2,
  \label{h and v_a n1}
          \end{gather}
with
          \begin{gather}
   s_1 = q^{ a (N_1 + N_2) / 2 } a_{1+} +
   {e} ^{ {i} \phi_r / 2 }  {1 \over
  \left[ k-1 \right]_q!} (a_{1-})^{k-1},
  \label{h and v_a n2-1}
  \\
   s_2 = a_{2-}  q^{- a (N_1 - N_2) / 2 }+
   {e} ^{ {i} \phi_r / 2 }  {1 \over
  \left[ k-1 \right]_q!} (a_{2+})^{k-1},
  \label{h and v_a n2}
          \end{gather}
where $a$ and $\phi_r$ are two real parameters. The parameter
$\phi_r$ is taken in the form
          \[
 \phi_r = \pi (k-1) r, \qquad r \in \mathbb{R}.
          \]
It is immediate to show that the action of $h$ and $v_{ra}$ on
${\cal F}_k$ is given by
          \begin{gather}
  h |n_1 , n_2) = {\sqrt{ n_1 (n_2 + 1) } |n_1 , n_2)},
  \qquad n_i = 0, 1, 2, \dots, k-1,
  \quad i=1,2
  \label{action de h sur n1n2}
          \end{gather}
and
          \begin{gather}
  v_{ra} |n_1 , n_2) = q^{a n_2} |n_1+1 , n_2-1),
  \qquad n_1 \not = k-1,
  \quad n_2 \not = 0,
  \label{action1 de v_a sur n1n2}
          \\
  v_{ra} |k-1 , n_2) =  {e} ^{ {i} {\phi}_r / 2 }
  q^{- a (k - 1 - n_2) / 2}
  |0 , n_2 - 1),
  \qquad n_2 \not= 0,
          \nonumber\\
  v_{ra} |n_1 , 0) =  {e} ^{ {i} {\phi}_r / 2 }
  q^{ a (k + n_1) / 2 }
  |n_1 + 1 , k-1),
  \qquad n_1 \not= k-1,\nonumber\\
  v_{ra} |k-1 , 0) =  {e} ^{{i} {\phi}_r}
  |0 , k-1).
  \label{action2 de v_a sur n1n2}
          \end{gather}

The operators $h$ and $v_{ra}$ satisfy interesting properties.
First, it is obvious that the operator~$h$ is Hermitian. Second,
the operator~$v_{ra}$ is unitary. In addition, the action of
$v_{ra}$ on the space ${\cal F}_k$ is cyclic. More precisely, we
can check that
          \begin{gather}
(v_{ra})^k =  {e} ^{ {i} \pi (k-1) (a + r) } I, \label{puissance k
de v_a}
          \end{gather}
where $I$ is the identity operator.

From the Schwinger work on angular momentum \cite{10Schwinger}, we
introduce
          \[
  J = {1 \over 2} \left( n_1+n_2 \right),  \qquad
  M = {1 \over 2} \left( n_1-n_2 \right).
          \]
Consequently, we can write
          \[
  |n_1 , n_2) = |J + M , J-M).
          \]
We shall use the notation
          \[
  |J , M \rangle := |J + M , J-M)
          \]
for the vector $|J + M , J-M)$. For a f\/ixed value of $J$, the
label $M$ can take $2J+1$ values $M = J, J-1, \dots, -J$.

For f\/ixed $k$, the following value of $J$
          \[
j = \frac{1}{2} (k-1)
          \]
is admissible. For a given value of $k \in \mathbb{N} \setminus
\{0 , 1\}$, the $k = 2j+1$ vectors $|j , m \rangle$ belong to the
vector space ${\cal F}_k$. Let $\varepsilon (j)$ be the subspace
of ${\cal F}_k$, of dimension $k$, spanned by the $k$ vectors $|j
, m \rangle$. We can thus associate $\varepsilon (j)$ for $j =
1/2, 1, 3/2, \dots$ with the values $k = 2,   3, 4,   \dots$,
respectively. We shall denote as $S$ the spherical basis
          \begin{gather}
S = \{ |j , m \rangle : m = j , j-1, \dots, -j \} \label{spherical
basis}
          \end{gather}
of the space $\varepsilon (j)$. The rewriting of
equations~(\ref{action de h sur n1n2})--(\ref{action2 de v_a sur
n1n2}) in terms of the vectors $|j , m \rangle$ shows that
$\varepsilon (j)$ is stable under $h$ and $v_{ra}$. More
precisely, we have the following result.

\begin{proposition}
The action of the operators $h$ and $v_{ra}$ on the
    subspace $\varepsilon (j)$ of ${\cal F}_k$ can be described by
          \[
  h |j , m \rangle = {\sqrt{ (j+m)(j-m+1) }} |j , m \rangle
          \]
and
          \[
  v_{ra} |j , m \rangle = \left( 1 - \delta_{m,j} \right) q^{(j-m)a}
  |j , m+1 \rangle + \delta_{m,j}
  {e}^{{i} 2 \pi j r}
  |j , -j \rangle,
          \]
which are a simple rewriting, in terms of the vectors $|j , m
\rangle$, of equations~\eqref{action  de h   sur n1n2},
        \eqref{action1 de v_a sur n1n2} and
        \eqref{action2 de v_a sur n1n2}, respectively.
\end{proposition}

We can check that the operator $h$ is Hermitian and the operator
$v_{ra}$ is unitary on the space~$\varepsilon (j)$.
Equation~(\ref{puissance k de v_a}) can be rewritten as
          \begin{gather}
  \left( v_{ra} \right)^{2j+1} =  {e} ^{ {i} 2{\pi} j (a + r) } I,
  \label{cyclic}
          \end{gather}
which ref\/lects the cyclic character of $v_{ra}$ on $\varepsilon
(j)$.

\subsection[The su$_2$ algebra]{The su$\boldsymbol{{}_2}$ algebra}

We are now in a position to give a realization of the Lie  algebra
su$_2$ of the group SU$_2$ in terms of the generators of $A_1$ and
$A_2$.  Let us def\/ine the three operators
          \begin{gather}
  j_+ = h           v_{ra},  \qquad
  j_- = v_{ra}^{\dagger} h,  \qquad
  j_z = \frac{1}{2} ( h^2 - v_{ra}^{\dagger} h^2 v_{ra} ).
  \label{definition of the j's}
          \end{gather}
It is straightforward  to check that the action on the vector $ |
j , m \rangle $ of the operators def\/ined by
equation~(\ref{definition of the j's}) is given by
          \begin{gather}
  j_+ |j , m \rangle    =  q^{+(j - m + s - 1/2)a}
  {\sqrt{ (j - m)(j + m+1) }}
  |j , m + 1 \rangle,
  \label{action des j sur jm signe plus}
  \\
  j_- |j , m \rangle    =  q^{-(j + m + s + 1/2)a}
  {\sqrt{ (j + m)(j - m+1) }}
  |j , m - 1 \rangle
  \label{action des j sur jm}
          \end{gather}
and
          \[
  j_z   |j , m \rangle = m |j,m \rangle.
          \]
Consequently, we get the following result.

\begin{proposition}
We have the commutation relations
          \begin{gather}
  \left[ j_z,j_{+} \right] = + j_{+},  \qquad
  \left[ j_z,j_{-} \right] = - j_{-},  \qquad
  \left[ j_+,j_- \right] = 2j_z,
  \label{adL su2}
          \end{gather}
which correspond to the Lie algebra {\rm su$_2$}.
\end{proposition}

We observe that the latter result (\ref{adL su2}) does not depend
on the parameters $a$ and $r$. On the contrary, the action of
$j_{\pm}$ on $| j , m \rangle$ depends on $a$ but not on $r$; the
familiar Condon and Shortley phase convention used in atomic
spectroscopy amounts to take $a = 0$ in equations~(\ref{action des
j sur jm signe plus}) and~(\ref{action des j sur jm}). The
decomposition (\ref{definition of the j's}) of the Cartan operator
$j_z$ and the shift operators~$j_+$ and~$j_-$ in terms of $h$ and
$v_{ra}$ constitutes a two-parameter polar decomposition of
su$_2$. Note that one-parameter polar decompositions were obtained
(i) in \cite{11JMLL} for SU$_2$ in a completely dif\/ferent
context and (ii) in \cite{12Ellinas-1, 12Ellinas-2, 12Ellinas-3,
12Ellinas-4} for SU$_n$ in the context of deformations.

\subsection[The W$_\infty$ algebra]{The W$\boldsymbol{{}_\infty}$ algebra}

In the following, we shall restrict $a$ to take the values $a = 0,
1, \dots, 2j$. By def\/ining the linear operator $z$ through
          \begin{gather}
  z   |j , m \rangle  =  q^{j-m} |j,m \rangle,
  \label{definition of z}
          \end{gather}
we can rewrite $v_{ra}$ as
          \begin{gather}
  v_{ra} = v_{r0} z^a, \qquad a = 0, 1, \dots, 2j.
  \label{vra en fonction de vr0}
          \end{gather}
The operators $v_{ra}$ and $z$ satisfy the $q$-commutation
relation
          \[
v_{ra} z - q z v_{ra} = 0.
          \]
Let us now introduce
          \[
t_m = q^{- m_1 m_2 / 2 } (v_{ra})^{m_1} z^{m_2}, \qquad m = (m_1 ,
m_2) \in {\mathbb{N}^{\ast}}^2.
          \]
Then, it is easy to obtain the following result.

\begin{proposition}
We have the commutator
          \[
[t_m , t_n] = 2 {i} \sin \left( \frac{\pi}{2j+1} m \wedge n
\right) t_{m+n},
          \]
where
          \begin{gather*}
 m = (m_1 , m_2) \in {\mathbb{N}^{\ast}}^2,
 \qquad  n = (n_1 , n_2) \in {\mathbb{N}^{\ast}}^2,
          \\
 m \wedge n = m_1 n_2 - m_2 n_1,     \qquad  m + n = (m_1 + n_1 , m_2 + n_2),
          \end{gather*}
so that the linear operators $t_m$ span the infinite-dimensional
Lie algebra {\rm W$_\infty$} introduced in~{\rm \cite{13FFZ}}.
\end{proposition}

This result parallels the ones obtained, on one hand, from a study
of $k$-fermions and of the Dirac quantum phase operator through a
$q$-deformation of the harmonic oscillator \cite{8DaoHasKib-1,
8DaoHasKib-2, 8DaoHasKib-3} and, on the other hand, from an
investigation of correlation measure for f\/inite quantum systems
\cite{12Ellinas-1, 12Ellinas-2, 12Ellinas-3, 12Ellinas-4}.

To close this section, we note that the (Weyl--Pauli) operators
$z$ and $v_{ra}$ can be used to generate the (Pauli) group ${\cal
P}_{2j+1}$ introduced in \cite{14PateraZassenhaus} (see
also~\cite{14bisWeyl}). The group ${\cal P}_{2j+1}$ is a f\/inite
subgroup of GL($2j+1, \mathbb{C}$) and consists of generalized
Pauli matrices. It is spanned by two generators. In fact, the
$(2j+1)^3$ elements of ${\cal P}_{2j+1}$ can be generated by
$v_{00}$ ($r=0$, $a=0$) and~$z$ for $2j+1$ odd and by $v_{10}$
($r=1$, $a=0$) and $ {e} ^{{i} \pi/(2j+1)}z$ for $2j+1$ even.

\section[An alternative basis for the representation of SU$_2$]{An alternative basis for the representation
of SU$\boldsymbol{{}_2}$}

\subsection[The $\{ j^2 , v_{ra} \}$ scheme]{The $\boldsymbol{\{ j^2 , v_{ra} \}}$ scheme}

It is immediate to check that the Casimir operator $j^2$ of su$_2$
can be written as
          \[
j^2 = h^2 + j_z^2 - j_z = v_{ra} ^{\dagger} h^2 v_{ra} + j_z^2 +
j_z
          \]
or
          \begin{gather}
j^2 = \frac{1}{4} (N_1 + N_2) (N_1 + N_2 + 2). \label{j2 in terms
of N_i's}
          \end{gather}
Thus, the operators $j^2$ and $v_{ra}$ can be expressed in terms
of the generators of $A_1$ and $A_2$, see equations~(\ref{h and
v_a n1})--(\ref{h and v_a n2}), and
     (\ref{j2 in terms of N_i's}). It is a simple matter of calculation
to prove that $j^2$ commutes with $v_{ra}$ for any value of $a$
and $r$. Therefore, for f\/ixed $a$ and $r$, the commuting set $\{
j^2, v_{ra}\}$ provides us with an alternative to the familiar
commuting set $\{ j^2, j_z \}$ of angular momentum theory.

\subsection{Eigenvalues and eigenvectors}
The eigenvalues and the common eigenvectors of the complete set of
commuting operators $\{ j^2, v_{ra} \}$ can be easily found by
using standard techniques. This leads to the following result.

\begin{proposition}
The spectra of the operators $v_{ra}$ and $j^2$ are given by
          \begin{gather}
  v_{ra} | j \alpha ; r a \rangle  =  q^{j(a+r) - \alpha}
      | j \alpha ; r a \rangle, \qquad
  j^2 | j \alpha ; r a \rangle  =  j(j+1)
      | j \alpha ; r a \rangle,
  \label{eigenvalues}
          \end{gather}
where
          \begin{gather}
|j \alpha ; r a \rangle = \frac{1}{\sqrt{2j+1}} \sum_{m = -j}^{j}
q^{(j + m)(j - m + 1)a / 2 - j m r + (j + m)\alpha} | j , m
\rangle, \qquad \alpha = 0, 1, \dots, 2j \label{j alpha r a in
terms of jm}
          \end{gather}
and $q$ is given by \eqref{definition of q} with $k = 2j+1$. The
spectrum of $v_{ra}$ is nondegenerate. For f\/ixed $j$, $a$, and
$r$, the $2j+1$ eigenvectors $|j \alpha ; r a \rangle$, with
$\alpha = 0, 1, \dots, 2j$, of the operator $v_{ra}$ generate an
orthonormalized basis $B_{ra} = \{ |j \alpha ; r a \rangle :
\alpha = 0, 1, \dots, 2j \}$ of the space $\epsilon(j)$. In
addition, we have
          \begin{gather}
\vert \langle j , m | j \alpha ; r a  \rangle \vert = {1 \over
\sqrt{2 j + 1}}, \qquad m = j, j-1, \dots, -j, \quad \alpha = 0,
1, \dots, 2j, \label{MUB n1}
          \end{gather}
so that the bases $B_{ra}$ and $S$ are mutually unbiased.
\end{proposition}

Let us recall that two orthonormalized bases $\{ |a \alpha \rangle
: \alpha = 0, 1, \dots, d-1 \}$ and $\{ |b \beta  \rangle : \beta
= 0, 1, \dots, d-1 \}$ of a $d$-dimensional Hilbert space over
$\mathbb{C}$, with an inner product denoted as $\langle \, | \,
\rangle$, are said to be mutually unbiased \cite{15Delsarte,
16Ivanovic, 17lesWootters-1, 17lesWootters-2, 17lesWootters-3,
17lesWootters-4, 17lesWootters-5, 17lesWootters-6, 18Calderbank}
if and only if
          \begin{gather}
| \langle a \alpha | b \beta \rangle | = \delta_{      a ,  b    }
\delta_{ \alpha , \beta } + (1 - \delta_{      a ,  b    })
\frac{1}{\sqrt{d}}. \label{MUB n2}
          \end{gather}
The correspondence between equations~(\ref{MUB n1}) and (\ref{MUB
n2}) is as follows. In equation~(\ref{MUB n1}),   $2j+1$
corresponds to $d$ while the symbols $jra$, $\alpha$, $j$, and $m$
correspond  to the symbols $a$, $\alpha$, $b$, and $\beta$,
respectively.

\subsection[Representation of SU$_2$]{Representation of SU$\boldsymbol{{}_2}$}

The representation theory of SU$_2$ can be transcribed in the $\{
j^2 , v_{ra} \}$ scheme. For f\/ixed $a$ and~$r$, the nonstandard
basis $B_{ra}$ turns out to be an alternative to the standard or
spherical basis $S$. In the $\{ j^2 , v_{ra} \}$ scheme, the
rotation matrix elements for the rotation $R$ of SO$_3$ assumes
the form
          \begin{gather}
D^{(j)}(R)_{\alpha \alpha'} = \frac{1}{2j+1} \sum_{m  = -j}^{j}
\sum_{m' = -j}^{j} q^{-\rho(j , m  , a, r, \alpha) +
    \rho(j , m' , a, r, \alpha')} \>
{\cal D}^{(j)}(R)_{mm'} \label{D for j2var in terms of D for j2jz}
          \end{gather}
in terms of the standard matrix elements ${\cal D}^{(j)}(R)_{mm'}$
corresponding to the $\{ j^2 , j_z \}$ scheme. In equation~(\ref{D
for j2var in terms of D for j2jz}), the function $\rho$ is
def\/ined by
          \begin{gather}
   \rho(J, M, x, y, z) = \frac{1}{2}(J + M)(J - M + 1)x - J M y + (J + M)z.
   \label{definition of rho}
          \end{gather}
Then, the behavior of the vector $|j \alpha ; r a \rangle$ under
an arbitrary rotation $R$ is given by
          \begin{gather}
P_{R} |j \alpha ; r a \rangle = \sum_{\alpha' = 0}^{2j} |j \alpha'
; r a \rangle  \> D^{(j)}(R)_{\alpha' \alpha}, \label{P_R sur j
alpha a r}
          \end{gather}
where $P_{R}$ stands for the operator associated with $R$. In the
case where $R$ is a rotation around the $z$-axis,
equation~(\ref{P_R sur j alpha a r}) takes a simple form as shown
in the following result.

\begin{proposition}
If $R(\varphi)$ is a rotation of an angle
          \begin{gather}
\varphi = p \frac{2 \pi}{2j+1}, \qquad p = 0, 1, 2, \dots, 2j
\label{Result 6-1}
          \end{gather}
around the $z$-axis, then we have
          \begin{gather}
P_{R(\varphi)} \> |j \alpha  ; r a \rangle = q^{jp} |j \alpha' ; r
a \rangle, \quad \alpha' = \alpha - p \ ({\rm mod} \ 2j+1),
\label{Result 6-2}
          \end{gather}
so that the set $\{ |j \alpha  ; r a \rangle : \alpha \ {\rm
ranging} \}$ is stable under $P_{R(\varphi)}$.
\end{proposition}

Consequently, the set $\{ |j \alpha ; r a \rangle : \alpha = 0, 1,
\dots, 2j \}$ spans a reducible representation of dimension $2j+1$
of the cyclic subgroup $C_{2j+1}$ of SO$_3$. It can be seen that
this representation is nothing but the regular representation of
$C_{2j+1}$. Thus, this representation contains each irreducible
representation of $C_{2j+1}$ once and only once. The nonstandard
basis $B_{ra}$ presents some characteristics of a group-subgroup
type basis in the sense that $B_{ra}$ carries a representation of
a subgroup of SO$_3$. However, this representation is reducible
(except for $j=0$). Therefore, the label $ \alpha ; a r$ does not
correspond to some irreducible representation of a subgroup of
SU$_2$ or ${\rm SO}(3) \sim {\rm SU}_2/Z_2$ so that the basis
$B_{ra}$ also exhibits some characteristics of a nongroup-subgroup
type basis.

\subsection[Wigner-Racah algebra of SU$_2$]{Wigner--Racah algebra of SU$\boldsymbol{{}_2}$}

We are now ready to give the starting point for a study of the
Wigner--Racah algebra of SU$_2$ in the $\{ j^2 , v_{ra} \}$
scheme. In such a scheme, the coupling or Clebsch--Gordan
coef\/f\/icients read
          \begin{gather}
  \left( j_1 j_2 \alpha_1 \alpha_2 |j_3 \alpha_3 \right)_{ra} =
  \left[         (2j_1 + 1)
                 (2j_2 + 1)
         (2j_3 + 1) \right]^{-\frac{1}{2}}
         \sum_{m_1=-j_1}^{j_1}
  \sum_{m_2=-j_2}^{j_2}
  \sum_{m_3=-j_3}^{j_3}\!
  ( j_1 j_2 m_1 m_2 | j_3 m_3 )
  \nonumber
  \\
\phantom{\left( j_1 j_2 \alpha_1 \alpha_2 |j_3 \alpha_3
\right)_{ra} =}{}\times
         q_1^{- \rho(j_1 , m_1 , a, r, \alpha_1)}     \>
         q_2^{- \rho(j_2 , m_2 , a, r, \alpha_2)}     \>
     q_3^{  \rho(j_3 , m_3 , a, r, \alpha_3)},
  \label{coupling coefficient}
          \end{gather}
where the function $\rho$ is given by equation~(\ref{definition of
rho}). In equation~(\ref{coupling coefficient}), $( j_1 j_2 m_1
m_2 | j_3 m_3 )$ is a~standard Clebsch--Gordan coef\/f\/icient in
the $\{ j^2 , j_z \}$ scheme and
          \[
q_\ell = \exp \left( {i} {2 \pi \over 2 j_\ell + 1 } \right),
\qquad \ell = 1,2,3.
          \]
The algebra of the new coupling coef\/f\/icients (\ref{coupling
coefficient}) can be developed in a way similar to the
one~\cite{19Edmonds} known in the $\{ j^2 , j_z \}$ scheme (see
\cite{9KibDao-1} for the basic ideas). In particular, following
the technique developed in \cite{3Kibler-1, 3Kibler-2}, the
familiar 6-$j$ and 9-$j$ symbols of Wigner can be expressed in
terms of the coupling coef\/f\/icients def\/ined by
equation~(\ref{coupling coefficient}).

\subsection[Realization of $v_{ra}$]{Realization of $\boldsymbol{v_{ra}}$}

The operator $v_{ra}$ can be expressed in the enveloping algebra
of SU$_2$. A possible way to f\/ind a realization of $v_{ra}$ in
terms of the generators $j_\pm$ and $j_z$ of SU$_2$ is as follows.

The f\/irst step is to develop $v_{ra}$ on the basis of the Racah
unit tensor ${\bf u}^{(k)}$ \cite{20Racah}. Let us recall that the
components ${u}^{(k)}_{p}$ of ${\bf u}^{(k)}$, $p = k, k-1, \dots,
-k$, are def\/ined by
          \[
 \langle j , m | u^{(k)}_{p} | j , m' \rangle = (-1)^{j-m}
 \begin{pmatrix}
 j     &k    &j    \cr
 -m    &p    &m'   \cr
 \end{pmatrix},
          \]
where the symbol ($\cdots$) stands for a 3-$jm$ Wigner symbol. The
Hilbert--Schmidt scalar product of ${u}^{(k)}_{p}$ by
${u}^{(\ell)}_{q}$ satisf\/ies
          \begin{gather}
 {\rm Tr}_{\epsilon(j)} \left( ({u}^{(k)}_{p})^{\dagger} {u}^{(\ell)}_{q} \right) =
 \Delta(j,j,k) \> \delta_{k,\ell} \> \delta_{p,q} \> \frac{1}{2k+1},
 \label{trace de uu}
          \end{gather}
where $\Delta(j,j,k) = 1$ if $j$, $j$, and $k$ satisfy the
triangular inequality and is zero otherwise. Therefore, the
coef\/f\/icients $b_{kq}(ra)$ of the development
          \begin{gather}
  v_{ra} = \sum_{k=0}^{2j} \sum_{p=-k}^{k} b_{kp}(ra) {u}^{(k)}_{p}
  \label{development of v}
          \end{gather}
can be easily calculated from equations~(\ref{trace de uu}) and
(\ref{development of v}). This yields
          \begin{gather}
  b_{kp}(ra) = (2k+1)
  {\rm Tr}_{\epsilon(j)} \left( ({u}^{(k)}_{p})^{\dagger} v_{ra} \right).
  \label{trace de uv}
          \end{gather}
By developing the rhs of (\ref{trace de uv}), it is possible to
obtain
          \begin{gather}
  b_{kp}(ra) = \delta_{p,1} \>
  (2k+1) \sum_{m=-j}^{j-1} q^{(j-m)a} (-1)^{j-m-1}
  \begin{pmatrix}
  j       &k    &j    \cr
  -m-1    &1    &m    \cr
  \end{pmatrix}
  \nonumber
  \\
  \phantom{b_{kp}(ra) =}{}+ \delta_{k,2j} \> \delta_{p,-2j} \>
  \sqrt{4 j + 1}  {e} ^{ {i} 2 \pi j r }.
  \label{expresssion de b}
          \end{gather}

The second step is to express ${u}^{(k)}_{p}$ in the enveloping
algebra of SU$_2$. This can be achieved by using the formulas
given in \cite{21GreKib-1, 21GreKib-2}. Indeed, the operator
${u}^{(k)}_{p}$ acting on $\varepsilon(j)$ reads
          \begin{gather}
u_{p}^{(k)} = \left[ \frac{(k-p)!}{(k+p)!(2j-k)!(2j+k+1)!}
\right]^{1/2} (-1)^{k+p} j_{+}^{p} \left[ {
(-1)^{p}\frac{(2j-p)!(k+p)!}{p!(k-p)!} } \right. \nonumber
\\
\phantom{u_{p}^{(k)} =}{} + (1-\delta_{k,p})
      \sum_{z=p+1}^{k} \left. { (-1)^{z}\frac{(2j-z)!(k+z)!} {z!(k-z)!(z-p)!}
      \prod_{t=1}^{z-p} (j + j_z + p - z + t) } \right]
\label{ukq, q positif}
          \end{gather}
for $p \geq 0$. The formula for $p<0$ may be derived from
(\ref{ukq, q positif}) with the help of the Hermitian conjugation
property $u_{-p}^{(k)}=(-1)^{p}(u_{p}^{(k)})^{\dagger}$.
Alternatively, $u_{p}^{(k)}$ with $p<0$ may be obtained by
changing $p$, $j_{+}$ and $j_z$ into $-p$, $-j_{-}$ and $-j_z$
respectively, in the rhs of~(\ref{ukq, q positif}) and by
multiplying the expression so-obtained by $(-1)^{k+p}$. This
yields
          \begin{gather}
u_{p}^{(k)} = \left[
\frac{(k+p)!}{(k-p)!(2j-k)!(2j+k+1)!}\right]^{1/2} (-1)^{p}
j_{-}^{-p} \left[ { (-1)^{p}\frac{(2j+p)!(k-p)!}{(-p)!(k+p)!} }
\right. \nonumber
\\
\phantom{u_{p}^{(k)} =}{} + (1-\delta_{k,-p}) \sum_{z=-p+1}^{k}
\left. { (-1)^{z}\frac{(2j-z)!(k+z)!}{z!(k-z)!(z+p)!}
\prod_{t=1}^{z-p}(j- j_z -p-z+t) } \right] \label{ukq, q negatif}
          \end{gather}
for $p \leq 0$.

As a conclusion we have the following result.

\begin{proposition}
The development of $v_{ra}$ in the enveloping algebra of {\rm
SU$_2$} is given by \eqref{development of v}--\eqref{ukq, q
negatif}.
\end{proposition}

By way of illustration, let us consider the case $j = 1/2$, $1$,
and $3/2$ for $r=0$ and $a=0$.

{\bf Case} $j = 1/2$
          \[
v_{00} = \sqrt{3} \left( {u}^{(1)}_{-1} - {u}^{(1)}_{1} \right)
\Rightarrow v_{00} = j_+ + j_-.
          \]

{\bf Case} $j = 1$
          \[
v_{00} = \sqrt{5} {u}^{(2)}_{-2}
       - \sqrt{6} {u}^{(1)}_{ 1} \Rightarrow
v_{00} = \frac{1}{\sqrt{2}} j_+ + \frac{1}{2} (j_-)^2.
          \]

{\bf Case} $j = 3/2$
          \begin{gather*}
v_{00} = - \left( 1 + \sqrt{3} \right) \sqrt{\frac{6}{5}}
{u}^{(1)}_{1}
         + {\sqrt{7}} {u}^{(3)}_{-3}
     + \left( \sqrt{3} - 2 \right) \sqrt{\frac{7}{5}} {u}^{(3)}_{1}
       \nonumber
       \\
       \Rightarrow \quad
v_{00} = \frac{1}{\sqrt{3}} j_+ +
\left( \frac{1}{\sqrt{3}} -  \frac{1}{2} \right) j_+ \left( j_z +
\frac{3}{2} \right) \left( j_z - \frac{1}{2} \right)
 + \frac{1}{6} (j_-)^3.
       \nonumber
          \nonumber
      \end{gather*}

A program in MAPLE was run to get $v_{0a}$ for higher values of
$j$. The results were returned under the form
          \[
j_{+}\times ({\rm a \ polynomial \ in} \ j_{z}) + \frac{1}{(2j)!}
(j_{-})^{2j}.
          \]
For $a=0$ as above, one can remark and prove that the polynomial
in $j_{z}$\ is of degree $\leq 2\left\lceil j\right\rceil -2$ and
that its coef\/f\/icients are elements of the f\/ield $\mathbb{Q}[
\sqrt{2},\sqrt{3},\dots ]$ which form a set of vectors
over~$\mathbb{Q}$ of rank $\leq \left\lceil j\right\rceil $. Since
these two upper bounds are reached for most of the values of $j$
for which $v_{00}$ was computed, one expects in general
$\left\lceil j\right\rceil -2$ independent linear relations
between the coef\/f\/icients. These relations seemingly exhibit
some regularities.

\subsection[Connection between $B_{ra}$ and $B_{sb}$]{Connection between $\boldsymbol{B_{ra}}$ and
$\boldsymbol{B_{sb}}$}

The operators $v_{ra}$ and $v_{sb}$ do not commute in general. For
example, in the situation where $a=b=0$, a necessary and
suf\/f\/icient condition to have $[v_{r0} , v_{s0}] = 0$ is
          \[
js = jr + t, \qquad t \in \mathbb{Z}.
          \]
Going back to the general case $a \not= b$, the operators $v_{ra}$
and $v_{sb}$ satisfy the property
          \begin{gather}
{\rm Tr}_{\varepsilon(j)} \left( v_{ra} ^{\dagger} v_{sb} \right)
= \delta_{a,b} (2j+1) +
 {e} ^{ {i} ( \phi_s - \phi_r ) } - 1.
\label{trace}
          \end{gather}
A simple development of the rhs of equation~(\ref{trace}) and the
use of equation~(\ref{eigenvalues}) lead to
          \begin{gather}
\sum_{\alpha=0}^{2j} \sum_{\beta=0}^{2j} q^{\alpha - \beta} \vert
\langle j \alpha ; ra | j \beta ; sb \rangle \vert^2 =
\delta_{a,b} q^{j(r-s)} (2j+1) + q^{j(a + r - b - s)} [ {e} ^{ {i}
( \phi_s - \phi_r ) } - 1]. \label{sum rule rasb}
          \end{gather}

Let us now consider the overlap between two bases $B_{ra}$ and
$B_{sb}$ corresponding to the schemes $\{ j^2, v_{ra} \}$ and
                             $\{ j^2, v_{sb} \}$, respectively. We have
          \begin{gather}
\langle j \alpha ; ra | j' \beta ; sb \rangle = \delta_{j,j'} \>
\frac{1}{2j+1} \> \sum_{m=-j}^j q^{\rho(j,m,b - a,s - r,\beta -
\alpha)}.
   \label{overlap rasb}
          \end{gather}
From equation~(\ref{overlap rasb}), we see that the overlap
$\langle j \alpha ; ra | j \beta ; sb \rangle$ depends solely on
the dif\/ference $\alpha - \beta$ rather than on $\alpha$ and
$\beta$ separately. Hence, equation~(\ref{sum rule rasb}) can be
reduced to
          \[
\sum_{\alpha =0}^{2j}q^{\alpha }\left\vert \left\langle j\alpha
;ra|j0;sb\right\rangle \right\vert ^{2}=\delta
_{a,b}q^{j(r-s)}+\frac{1}{2j+1} q^{j(a+r-b-s)} \left[  {e}
^{{i}(\phi_{s}-\phi_{r})}-1 \right],
          \]
a relation that also follows from repeated applications of
equations~(\ref{Result 6-1}) and
     (\ref{Result 6-2}) to the lhs of~(\ref{sum rule rasb}).

In the special case where $b=a$, we get
          \[
 \langle j \alpha ; ra | j' \beta ; sa \rangle = \delta_{j,j'} \>
\frac{1}{2j+1} \> q^{j(\beta - \alpha)} \> \frac{\sin \pi
(jr - \alpha - js + \beta)}
     {\sin \frac{\pi}{2j+1} (jr - \alpha - js + \beta)}
          \]
for $jr - \alpha - js + \beta \not\equiv 0$ (mod $2j+1$) and
          \[
\langle j \alpha ; ra | j' \beta ; sa \rangle = \delta_{j,j'} \>
(-1)^{2jk} \> q^{j(\beta - \alpha)}
          \]
for $jr - \alpha - js + \beta = (2j+1)k$ with $k \in \mathbb{Z}$.
It is clear that for $s=r$, we recover that the basis $B_{ra}$ is
orthonormalized since equation~(\ref{overlap rasb}) gives
          \begin{gather}
\langle j \alpha ; ra | j \beta ; ra \rangle =
\delta_{\alpha,\beta}. \label{orthonormality}
          \end{gather}

The case $b \not= a$ is much more involved. For $b \not= a$ and $r
= s$, equation~(\ref{overlap rasb}) is amenable in the form of a
generalized quadratic Gauss sum $S(u, v, w)$. Such a sum is
def\/ined by
          \begin{gather}
S(u, v, w) = \sum_{k=0}^{|w| -1}  {e} ^{ {i} \pi (uk^2 + vk) / w
},
   \label{65 generalized quadratic Gauss sum}
          \end{gather}
where $u$, $v$, and $w$ are integers such that $uw \not= 0$ and
$uw + v$ is an even integer \cite{22Berndt}. As a matter of fact,
we have the following result.

\begin{proposition}
For $b \not= a$, the overlap $\langle j \alpha ; ra | j \beta ; rb
\rangle$ can be written as
          \begin{gather}
\langle j \alpha ; ra | j \beta ; rb \rangle = \frac{1}{w} S(u, v,
w),
   \label{overlap en fonction de S}
          \end{gather}
where
          \[
u = a - b, \qquad v = -(a - b)(2j+1) - 2(\alpha - \beta), \qquad w
= 2j+1,
          \]
with $a-b           = \pm 1, \pm 2, \dots, \pm 2j$ and  $\alpha,
\beta = 0, 1, \dots, 2j$. Furthermore, for $2j+1$ prime we have
          \begin{gather}
\vert \langle j \alpha ; ra | j \beta ; rb \rangle \vert =
\frac{1}{\sqrt{2j+1}},
  \label{module de l'overlap pour 2j+1 premier}
          \end{gather}
with $a-b           = \pm 1, \pm 2, \dots, \pm 2j$ and  $\alpha,
\beta = 0, 1, \dots, 2j$.
\end{proposition}

The proof of Proposition 8 is given in Appendix~A. Along this
vein, relations between genera\-lized quadratic Gauss sums and the
absolute value of a particular Gaussian sum are presented in
Appendices~B and~C, respectively.

It is to be noted that equation~(\ref{module de l'overlap pour
2j+1 premier}) can be proved equally well without using
generalized quadratic Gauss sums. The following proof is an
adaptation, in the framework of angular momentum, of the method
developed in \cite{23Bandyo} (see also \cite{24Lawrence,
25Klimov-1, 25Klimov-2, 25Klimov-3}) in order to construct
a~complete set of MUBs in $\mathbb{C}^d$ with $d$ prime.

\begin{proof}
We start from
          \[
v_{ra} z^n = v_{rb}, \qquad b = a + n, \qquad n \in \mathbb{Z},
          \]
which can be derived from equation~(\ref{vra en fonction de vr0}).
In view of Proposition 5, the action of the operator $v_{ra} z^n$
on the vector $\vert j \beta_0 ; rb \rangle$ leads to
          \begin{gather}
v_{ra} z^n \vert j \beta_0 ; rb \rangle = q^{j(a + n + r) -
\beta_0} \vert j \beta_0 ; rb \rangle. \label{action de vra par
zn}
          \end{gather}
Furthermore, equations~(\ref{definition of z}) and (\ref{j alpha r
a in terms of jm}) give
          \begin{gather}
z^n \vert j \beta_0 ; rb \rangle = q^{2jn} \vert j \beta_1 ; rb
\rangle, \qquad \beta_i = \beta_0 - in, \qquad i \in \mathbb{Z},
\quad n \in \mathbb{Z}. \label{action de zn}
          \end{gather}
Let us consider the scalar product $\langle j \alpha ; ra \vert
v_{ra} z^n \vert j \beta_0 ; rb \rangle$. This product can be
calculated in two dif\/ferent ways owing to (\ref{action de vra
par zn}) and (\ref{action de zn}). We thus obtain
          \begin{gather}
\left\vert \langle j \alpha ; ra \vert v_{ra} \vert j \beta_1 ; rb
\rangle \right\vert = \left\vert \langle j \alpha ; ra \vert j
\beta_0 ; rb \rangle \right\vert.
 \label{module = module}
          \end{gather}
Since $v_{ra}$ is unitary and satisf\/ies (\ref{cyclic}), we can
write
          \begin{gather}
 v_{ra} =
 \left( v_{ra}^{\dagger} \right)^{2j} \left( v_{ra} \right)^{2j} v_{ra} =
 \left( v_{ra}^{\dagger} \right)^{2j} \left( v_{ra} \right)^{2j+1}      =
  {e} ^{{i} 2 \pi (a + r)j} \left( v_{ra}^{\dagger} \right)^{2j}.
 \label{magic}
          \end{gather}
Finally, the introduction of (\ref{magic}) into (\ref{module =
module}) produces the master formula
          \[
 \left\vert
 \langle j \alpha ; ra \vert j \beta_1 ; rb \rangle
\right\vert = \left\vert \langle j \alpha ; ra \vert j \beta_0 ;
rb \rangle \right\vert.
          \]
The number of dif\/ferent $\beta_i$ modulo $2j+1$ that can be
reached by repeated translations of $\beta_0$ is $(2j+1)/{\rm
gcd}(2j+1 , \vert n \vert)$. As a conclusion,
equation~(\ref{module de l'overlap pour 2j+1 premier}) is true for
$2j+1$ prime.
\end{proof}

\section{Applications to cyclic systems and quantum information}

We shall devote the rest of this paper to some applications
involving state vectors of type (\ref{j alpha r a in terms of
jm}). More specif\/ically, we shall deal with cyclic systems, like
ring shape molecules and $1/2$-spin chains, and with MUBs of
quantum information theory (quantum cryptography and quantum
tomography). For the sake of comparison with some previous works,
it is appropriate to leave the framework of angular momentum
theory by transforming sums on $m$ from $-j$ to $j$ into sums on
$k$ from $0$ to $d$ with $d=2j+1$. Furthermore, we shall limit
ourselves in Section 4 to vanishing $r$-parameters. Consequently,
we shall use the notation $|a \alpha \rangle := |j \alpha ; 0 a
\rangle$ and $| k \rangle := | j m \rangle$ with $k = j+m$. Thus,
equation~(\ref{j alpha r a in terms of jm}) becomes
          \begin{gather}
    |a \alpha \rangle = \frac{1}{\sqrt{d}} \sum_{k=0}^{d-1}
    q^{k(d - k)a / 2 + k\alpha} | k \rangle,
    \qquad      a = 0, 1, \dots, 2j,
    \quad \alpha = 0, 1, \dots, 2j.
    \label{a alpha in terms of k}
          \end{gather}
The notation $|a \alpha \rangle$ and $| k \rangle$ is especially
adapted to the study of cyclic systems and MUBs.

\subsection{Cyclic systems}

Let us consider a ring shape molecule with $N$ atoms (or
aggregates) at the vertices of a regular polygon with $N$ sides
($N=6$ for the benzen molecule C$_6$H$_6$). The atoms are labelled
by the integer $n$ with $n = 0, 1, \dots, N-1$. Hence, the cyclic
character of the ring shape molecule makes it possible to identify
the atom with the number $n$ to the one with the number $n+kN$
where $k \in \mathbb{Z}$ (the location of an atom is def\/ined
modulo $N$). Let $| \varphi_n \rangle$ be the atomic state vector,
or atomic orbital in quantum chemistry parlance, describing a
$\pi$-electron located in the neighboring of site $n$. From
symmetry considerations, the molecular state vector, or molecular
orbital, for the molecule reads \cite{26LeBellac}
          \[
| \kappa_s \rangle = \frac{1}{\sqrt{N}} \sum_{n = 0}^{N-1}
 {e} ^{{i} 2 \pi n s / N } | \varphi_n \rangle,
          \]
with $s = 0, 1, \dots, N-1$. As a result, the molecular orbital $|
\kappa_s \rangle$ assumes the same form, up to a~global phase
factor, as the state $| a \alpha \rangle$ given by
equation~(\ref{a alpha in terms of k}) with $a=0$ and $\alpha =
s$.

A similar result can be obtained for a one-dimensional chain of
$N$ $1/2$-spins (numbered with $n=0, 1, \dots, N-1$) used as a
modeling tool of a ferromagnetic substance. Here again, we have a
cyclical symmetry since the spins numbered $n=N$ and $n=0$ are
considered to be identical. The spin waves can then be described
by state vectors (see \cite{26LeBellac}) very similar to the ones
given by equation~(\ref{a alpha in terms of k}) with again $a=0$.

\subsection{Mutually unbiased bases}
The results in Section 3 can be applied to the derivation of MUBs.
Proposition 8 provides us with a method for deriving MUBs that
dif\/fers from the methods developed, used or discussed
in~\cite{15Delsarte}--\cite{18Calderbank} and in
                               \cite{27Chaturvedi}--\cite{38Hayashi}.

\subsubsection{$d$ arbitrary}
Let $V_a$ be the matrix of the operator $v_{0a}$ in the
computational basis
          \begin{gather}
S = \{ | k \rangle    : k = d-1, d-2, \dots, 0 \},
\label{numerical basis}
          \end{gather}
cf.~equation~(\ref{spherical basis}). This matrix can be expressed
in terms of the generators $E_{x,y}$ (with $x,y = 0, 1, \dots,
d-1$) of the unitary group U$_d$. The generator $E_{x,y}$ is
def\/ined by its matrix elements
          \[
\left( E_{x,y} \right)_{ij} = \delta_{i,x} \> \delta_{j,y}, \qquad
i,j = d-1, d-2, \dots, 0.
          \]
Therefore, we have
          \[
    V_a = E_{0,d-1} + \sum_{k=1}^{d-1} q^{(d-k)a} E_{k,k-1},
          \]
with $q = \exp ( {2 \pi {i} / d} )$, as far as we label the lines
and columns of $V_a$ according to the decreasing order $d-1, d-2,
\dots, 0$. The eigenvectors $\varphi(a \alpha)$ of matrix $V_a$
are expressible in terms of the column vectors $e_x$ (with $x = 0,
1, \dots, d-1$) def\/ined via
          \[
\left( e_x \right)_{i1} = \delta_{i , x}, \qquad i = d-1, d-2,
\dots, 0.
          \]
Indeed, we have
          \[
\varphi(a \alpha) = \frac{1}{\sqrt{d}} \sum_{k=0}^{d-1} q^{
k(d-k)a / 2 + k \alpha } e_{k}.
          \]
This can be summed up by the following result.

\begin{proposition}
We have the eigenvalue equation
          \[
V_a \varphi(a \alpha) = q^{(d-1)a / 2 - \alpha} \varphi(a \alpha),
\qquad \alpha = 0, 1, \dots, d-1.
          \]
Furthermore, the generalized Hadamard matrix
          \begin{gather}
H_a = \sum_{\alpha = 0}^{d-1} \sum_{k=0}^{d-1} q^{ k(d-k)a /2 + k
\alpha } E_{k , \alpha} \label{generalized Hadamard matrix}
          \end{gather}
satisf\/ies
          \[
H_a^{\dagger} H_a = dI, \qquad H_a^{\dagger} V_a H_a = q^{ (d-1)a
/ 2 } \> d \> \sum_{\alpha=0}^{d-1} q^{- \alpha}  E_{\alpha ,
\alpha}
          \]
and thus reduces the endomorphism associated with $V_a$.
\end{proposition}

\subsubsection[$d$ prime]{$\boldsymbol{d}$ prime}

We are now in a position to establish contact with MUBs. We know
that for a $d$-dimensional Hilbert space, with $d$ prime ($d=p$)
or a power of a prime ($d=p^e$), with $p$ prime and $e$ positive
integer greater than~1, there exists a complete set of $d+1$
pairwise MUBs \cite{15Delsarte, 16Ivanovic, 17lesWootters-1,
17lesWootters-2, 17lesWootters-3, 17lesWootters-4,
17lesWootters-5, 17lesWootters-6, 18Calderbank, 27Chaturvedi,
28Pittinger-1, 28Pittinger-2, 29SPR-1, 29SPR-2, 29SPR-2bis,
29SPR-3, 29SPR-4, 29SPR-5}.

For $d = p = 2j+1$ prime, the bases
          \[
B_{0a} = \{ | a \alpha \rangle := |j \alpha ; 0 a \rangle : \alpha
= 0, 1, \dots, p-1 \}, \qquad a = 0, 1, \dots, p-1
          \]
satisfy (\ref{orthonormality}) and (\ref{module de l'overlap pour
2j+1 premier}). Consequently, they constitute an incomplete set of
$p$ MUBs. On the other hand, the bases $S$ and $B_{0a}$, with
f\/ixed $a$, are mutually unbiased (see Proposition~5). Therefore,
we have the following result.

\begin{proposition}
For $p$ prime, the spherical or computational basis $S$, given by
\eqref{spherical basis} or \eqref{numerical basis}, and the bases
$B_{0a}$ with $a = 0, 1, \dots, p-1$, given by \eqref{j alpha r a
in terms of jm} or \eqref{a alpha in terms of k}, constitute a
complete set of $p+1$ MUBs in $\mathbb{C}^p$. In matrix form, the
basis vectors of each $B_{0a}$ are given by the columns of the
generalized Hadamard matrix \eqref{generalized Hadamard matrix}.
\end{proposition}

It is to be noted that each column of matrix $H_a$ gives one of
the $p$ vectors of basis $B_{0a}$. For~$p$ prime, the $(p+1)p$
vectors of the $p+1$ MUBs are given by the columns of the $p$
matrices $H_a$ with $a = 0, 1, \dots, p$ together with the columns
of the $p$-dimensional identity matrix. We observe that
Proposition 10 is valid for $p=2$.

\subsubsection[$d$ not prime]{$\boldsymbol{d}$ not prime}
Returning to the general case where $d$ is arbitrary, it is clear
that the bases $B_{0a}$ with $a = 0, 1, \dots, d-1$ do not
constitute in general a set of $d$ pairwise MUBs. However, some of
them can be mutually unbiased. An easy way to test the unbiased
character of the bases $B_{0a}$ and $B_{0b}$, with $a \not= b$, is
to calculate $H_a^{\dagger} H_b$; if the module of each matrix
element of $H_a^{\dagger} H_b$ is equal to~$\sqrt{d}$, then
$B_{0a}$ and $B_{0b}$ are mutually unbiased.

The bases $B_{0a}$ corresponding to $d = p = 2j+1$, $p$ prime, can
serve to generate MUBs in the case $d = p^e = (2j+1)^e$. For $d =
p^e = (2j+1)^e$, with $p$ prime and $e$ positive integer greater
than~1, let us consider the $p^e$ bases
          \[
B_{a_1 a_2 \cdots a_e} = B_{0a_1} \otimes
                         B_{0a_2} \otimes \cdots \otimes
             B_{0a_e}, \qquad
a_i \in \{ 0, 1, \dots, p-1 \}, \quad i = 1, 2, \dots, e.
          \]
In this case, we face a degeneracy problem. Therefore, the basis
vectors in $B_{a_1 a_2 \cdots a_e}$ must be reorganized (via
linear combinations) in order to form, with the $p^e$-dimensional
computational basis, a complete set of $p^e + 1$ MUBs in
$\mathbb{C}^d$ (cf.~the approaches of \cite{23Bandyo, 24Lawrence,
25Klimov-1, 25Klimov-2, 25Klimov-3}).

Up to this point, we have dealt with pairwise MUBs. We close this
section with a few remarks concerning the number of bases which
are unbiased with a given basis. In the proof of Proposition 8
given in Appendix A, one of the key arguments is that $u$ or
$u^{\prime}$ must be invertible modulo $2(2j+1)$, what is
immediately checked since $2j+1$ is prime. This argument cannot be
used when the dimension $d=2j+1$ is a power of a prime, $d=p^{e}$
($p$ prime and $e$ integer greater than 1). However, taking $p
\neq 2$, let us consider the bases $B_{ra}$ ($a=0, 1, \dots, d-1$)
whose vectors are given by (\ref{j alpha r a in terms of jm}),
with $j=(d-1)/2$. We remark that the number of bases $B_{ra}$ ($a$
ranging) that are unbiased with one of them is at least $\varphi
(p^{e}) = p^{e} - p^{e-1}$, a remark that is also valid for
arbitrary dimension. If $d=p_{1}^{e_{1}}
   p_{2}^{e_{2}} \cdots
   p_{n}^{e_{n}}$, with $p_{i} \neq 2$ for $i = 1, 2, \dots, n$,
then the number of bases $B_{ra}$ ($a$ ranging) that are unbiased
with one of them is at least
          \[
\varphi(d) = \prod_{i=1}^{n}
              p_{i}^{e_{i}} - p_{i}^{e_{i}-1}.
          \]
These considerations can be expressed in a geometrical way in the
case of a prime power dimension $d=p^{e}$, with $p \neq 2$. Any
integer $a$ between $0$ and $p^{e}-1$ can be written in the form
          \[
a=a_{0}+a_{1}p+\cdots +a_{e-1}p^{e-1},
          \]
with $0 \leq a_{i} \leq p-1$ for $i = 0, 1, \dots, e-1$. Thus, any
basis $B_{ra}$ corresponds to the point of coordinates ($a_{0},
a_1, \dots, a_{e-1}$) in an af\/f\/ine space of dimension $e$\
over the Galois f\/ield $\mathbb{Z}/p\mathbb{Z}$. Moreover, two
bases $B_{ra}$\ and $B_{rb}$\ are mutually unbiased if and only if
$a_{0}-b_{0}\neq 0$, what excludes a hyperplane of the af\/f\/ine
space. Whenever $d$ is a product of prime powers, all of the
primes being dif\/ferent from $2$, a generalization is
straightforward by the use of the Chinese remainder theorem.

\section{Concluding remarks}

The originality of the present paper rests upon the development of
the representation of the nondeformed group SU$_2$ by means of
truncated deformed oscillators and of its application to cyclic
systems and MUBs. The main results of this work may be summarized
as follows.

From two deformed oscillator algebras, with a single deformation
parameter taken as a root of unity, we generated a new polar
decomposition of the nondeformed group SU$_2$. The latter
decomposition is characterized by a two-parameter operator
$v_{ra}$ and dif\/fers from previous decompositions. Our
decomposition proved to be especially adapted for generating the
inf\/inite-dimensional Lie algebra W$_\infty$. The set $\{
(v_{01})^n : n = 0, 1, \dots, 2j \}$ acting on a subspace
associated with the irreducible representation ($j$) of SU$_2$
spans the cyclic group $C_{2j+1}$. For f\/ixed $j$, each operator
$v_{0a}$ ($a = 0, 1, \dots, 2j$) corresponds to an irreducible
representation of $C_{2j+1}$. This yields a tower of
group-subgroup chains SO$_3 \supset C_{2j+1}$. The operator
$v_{ra}$ is a pseudo-invariant under~$C_{2j+1}$ and commutes with
the Casimir $j^2$ of SU$_2$. For an arbitrary irreducible
representation~($j$) of SU$_2$, we developed $v_{ra}$ in the
enveloping algebra of SU$_2$. Such a development might be useful
for a study of the Wigner--Racah agebra of SU$_2$ in a SU$_2
\supset C_{2j+1}^{\ast}$ basis. The eigenstates $|j \alpha ; ra
\rangle$ of the set $\{ j^2 , v_{ra} \}$ can be seen as
generalized discrete Fourier transforms (in the sense of
\cite{30Vourdas-1, 30Vourdas-2, 30Vourdas-3, 30Vourdas-4,
30Vourdas-5}) of the eigenstates of the standard set $\{ j^2 , j_z
\}$. This led us to strongly SABs $B_{ra} = \{ |j \alpha ; ra
\rangle : \alpha = 0, 1, \dots, 2j \}$ which are unbiased with the
spherical basis $S = \{ |j m \rangle : m = j, j-1, \dots, -j \}$.

The statevectors $|j \alpha ; ra \rangle$ are quite convenient to
study cyclic systems (as for example ring-shape molecules) and
MUBs. In particular, in the case where $2j+1$ is prime, we
generated from the statevectors $|j \alpha ; ra \rangle$ a
complete set of MUBs for the f\/inite-dimensional Hilbert space
$\mathbb{C}^{2j+1}$. The obtained MUBs are given by a single
formula which is easily codable on a computer. As a~by-product,
the close-form expression for the vectors $|j \alpha ; ra \rangle$
generates generalized Hadamard matrices in arbitrary dimension.
The new approach to MUBs developed in this work lies on the use of
(i) deformations introduced in fractional supersymmetric quantum
mechanics \cite{8DaoHasKib-1, 8DaoHasKib-2, 8DaoHasKib-3}, (ii)
angular momentum theory, and (iii) generalized quadratic Gauss
sums (for which we gave useful formulas in the appendices). In
this respect, it dif\/fers from the approaches developed in
previous studies through the use of Galois f\/ields and Galois
rings, discrete Wigner functions, mutually orthogonal Latin
squares, graph theory, and f\/inite geometries (e.g.,
see~\cite{31Klappenecker-1, 31Klappenecker-2, 32Grassl, 33Durt-1,
33Durt-2, 33Durt-3, 34Wocjan, 35Archer, 36Bengtsson-1,
36Bengtsson-2, 36Bengtsson-3, 36Bengtsson-4, 37Boykin} and
references cited therein for former works). Our approach to MUBs
in the framework of angular momentum should be particularly
appropriate for dealing with entanglement of spin states.

To close this article, let us say a few more words about point
(iii) just above. We carried on with the idea of building MUBs
using additive characters and Gauss sums \cite{17lesWootters-1,
17lesWootters-2, 17lesWootters-3, 17lesWootters-4,
17lesWootters-5, 17lesWootters-6, 27Chaturvedi, 29SPR-1, 29SPR-2,
29SPR-2bis, 29SPR-3, 29SPR-4, 29SPR-5}. However, our own formulas
come out with an extra 1/2 factor in the argument of the
exponential and we needed to study a special kind of Gauss sums to
prove mutual unbiasedness in prime dimension. These Gauss sums
exhibit two features of previous works \cite{17lesWootters-1,
17lesWootters-2, 17lesWootters-3, 17lesWootters-4,
17lesWootters-5, 17lesWootters-6,
31Klappenecker-1,31Klappenecker-2}, namely a second degree
polynomial as an argument of the exponential and the doubling of
the dimension in the denominator. Such combinations or comparisons
between dif\/ferent methods have been hardly explored up to now,
except in \cite{35Archer}.

When one tries to f\/igure out MUBs in a space the dimension of
which is a nontrivial power of a prime, $d=p^{e}$, $e>1$, one
usually refers to the additive characters of the Galois
f\/ield~$\mathbb{F}_{d}$. Therefore, the method is specif\/ic to
the case of prime power dimensions. In the end of Section~4.2, the
same Gauss sums as above enabled us to get some partial results
about unbiasedness for arbitrary dimensions. We f\/irst replaced
Galois f\/ields with the simplest Galois rings
$\mathbb{Z}/p^{e}\mathbb{Z}$ and then generalized the so-obtained
replacement through the use of the Chinese remainder theorem.
Unfortunately, numerical tests show that all bases obtained in
that way are not mutually unbiased, even in prime power dimension,
though other regularities appear. In order to study unbiasedness
in arbitrary dimension, it would be important to know whether or
not there exist a structure which supports both an equivalent of
the trace operator for Galois f\/ields and an equivalent of the
Chinese remainder theorem that could be combined ef\/f\/iciently.

Finally, it is also to be noted that in our geometrical
interpretation, bases are associated to points rather than to
striations of the space as in~\cite{17lesWootters-1,
17lesWootters-2, 17lesWootters-3, 17lesWootters-4,
17lesWootters-5, 17lesWootters-6} and \cite{38Hayashi}.

\appendix

\section{Proof of Proposition 8}

The proof of (\ref{overlap en fonction de S}) is straightforward.
Let us focus on the proof of (\ref{module de l'overlap pour 2j+1
premier}).

\begin{proof}
It is suf\/f\/icient to combine (\ref{overlap rasb}) for $j=j'$
and $s=r$ with (\ref{definition of rho}) and (\ref{65 generalized
quadratic Gauss sum}). This yields
          \[
(2j+1) \langle j \alpha ; ra | j \beta ; rb \rangle = S(u,v,2j+1)
= \sum_{k=0}^{2j}q^{(uk^{2}+vk)/2},
          \]
where $u=a-b$, $v = 2(\beta -\alpha )+(2j+1)(b-a)$, and $q= {e}
^{2 \pi {i} / (2j+1)}$.

For $j=1/2$, the generalized quadratic Gauss sum $S(u,v,2)$ can be
easily calculated and we then check that (\ref{module de l'overlap
pour 2j+1 premier}) is satisf\/ied for $2j+1=2$.

We continue with $2j+1$ equal to an odd prime number. In
$S(u,v,2j+1)$, the integer $u$ is such that $-2j \leq u \leq 2j$
and, for $2j+1$ prime with $j \not= 1/2$, the integer $v$ has the
same parity as $u$. We shall thus consider in turn $u$ even and
$u$ odd.

In the case $u$ even, $\xi=u/2$ and $\eta=v/2$ are two integers.
Then, we have
          \[
S(u,v,2j+1) = \sum_{k=0}^{2j}q^{\xi k^{2} + \eta k},
          \]
where the exponent of $q$ may be taken modulo $2j+1$. A
translation of the index $k$ gives
          \[
S(u,v,2j+1) = \sum_{k=0}^{2j}q^{\xi (k + t)^{2} + \eta (k+t)}.
          \]
By choosing $t$ such that $2 \xi t + \eta = 0$ (mod $2j+1$), we
get
          \begin{gather}
\vert S(u,v,2j+1) \vert = \left\vert \sum_{k=0}^{2j} q^{\xi k^{2}}
\right\vert. \label{module}
          \end{gather}
The value of the rhs of (\ref{module}) is $\sqrt{2j+1}$
\cite{22Berndt}. Therefore, (\ref{module de l'overlap pour 2j+1
premier}) is proved for $2j+1$ odd prime and $u$ even.

In the case $u$ odd, let us introduce the canonical additive
character of $\mathbb{Z}/(2(2j+1))\mathbb{Z}$
          \[
\psi : (\mathbb{Z}/(2(2j+1))\mathbb{Z}, +)
    \longrightarrow  (\mathbb{C}, \times) : x   \longmapsto      q^{y/2},
          \]
with $y \in \mathbb{Z}$ a representative of $x$ modulo $2(2j+1)$.
Consequently, we have
          \[
S(u,v,2j+1) = \sum_{k=0}^{2j} \psi (uk^{2}+vk),
          \]
where the argument of $\psi$ stands for a residue modulo
$2(2j+1)$. In order to apply the translation trick and to get rid
of the linear term, as in the even case, $k$ has to range over a
complete set of residues modulo $2(2j+1)$. For this purpose, we
may for instance consider the extra sum
          \begin{gather}
\sum_{\ell=2j+1}^{2(2j+1)-1} \psi (u\ell^{2} + v\ell) =
\sum_{k=0}^{2j} \psi (uk^{2} + 2(2j+1)uk + u(2j+1)^{2}+ vk +
v(2j+1)).
   \label{Extra sum}
          \end{gather}
The second term of the argument of $\psi$ in the rhs of
(\ref{Extra sum}) vanishes under $\psi$. Moreover,
          \begin{gather}
    u(2j+1)^{2} + v(2j+1) = 2(2j+1)uj + (u+v)(2j+1)
    \equiv 0 \  \ ({\rm mod} \ 2(2j+1))
    \label{96man}
          \end{gather}
since $u+v$ is even. Hence, the extra sum is equal to
$S(u,v,2j+1)$ so that
          \[
S(u,v,2j+1) = \frac{1}{2}\sum_{k=0}^{2(2j+1)-1} \psi (uk^{2}+vk).
          \]
Now let us carry out the translation
      \begin{gather}
u(k+t)^{2} + v(k+t) = uk^{2} + (2ut+v)k + ut^{2} + vt.
          \label{Translation}
      \end{gather}
Since $u$ is odd and between $-2j$ and $2j$, it is invertible
modulo $2(2j+1)$. Choosing $t \equiv u^{-1} \ ({\rm mod} \
2(2j+1))$, we see that
          \begin{gather}
\left\vert S(u, v  , 2j+1) \right\vert = \left\vert S(u, v+2,
2j+1) \right\vert,
    \label{modules des S}
          \end{gather}
where an increase of $v$ by $2$ amounts for an increase of $\beta
- \alpha$ by $1$. Therefore, the modules in the lhs of
(\ref{module de l'overlap pour 2j+1 premier}) do not depend on
$\beta - \alpha$. To show that they are independent of $a-b$, we
need only to remember that the overlaps $\left\langle j \alpha ;
ra | j \beta ; rb \right\rangle$ are coef\/f\/icients connecting
two orthonormalized bases. Consequently
          \[
\sum_{\alpha =0}^{2j}\left\vert \left\langle j\alpha ;ra|j\beta
;rb\right\rangle \right\vert ^{2} = 1
          \]
and
          \[
\forall \; \alpha = 0, 1, \dots, 2j : \ (2j+1) \left\vert
\left\langle j \alpha ; ra|j\beta ;rb\right\rangle \right\vert^{2}
= 1,
          \]
so that (\ref{module de l'overlap pour 2j+1 premier}) is proved
for $2j+1$ prime and $u$ odd.
\end{proof}

At this point it is interesting to emphasize that the method we
have developed to handle the odd case works in the even case too.
Suppose $u=2^{n}u^{\prime}$, with $u^{\prime}$ not divisible by
$2$. In the translation relation (\ref{Translation}), the term
$2ut$ should be replaced by $2^{n+1}u^{\prime}t$, where
$u^{\prime}$ is invertible modulo $2(2j+1)$. Thus $v+2$ in
(\ref{modules des S}) is replaced by $v + 2^{n+1}$ and an increase
of $v$ by $2^{n+1}$ amounts for an increase of $\beta - \alpha$ by
$2^{n}$. Since $2^{n}$ is coprime with $2j+1$, all values of
$\beta - \alpha$ will be swept over modulo $2j+1$ and the result
follows.

\section{Relations between generalized quadratic Gauss sums}

As a by-product of this work, it is worthwhile to mention that the
method of translation among a complete set of residues, recurrent
in the present paper, can be used to derive relations between
generalized quadratic Gauss sums. The Gauss sum $S(u,v,w)$, with
$u$, $v$, and $w$ integers such that $w \neq 0$ and $uw+v$ even,
see (\ref{65 generalized quadratic Gauss sum}), can be rewritten
as
          \begin{gather}
S(u,v,w) = q^{(ut^{2}+vt)/2} \sum_{k=-t}^{\vert w \vert - 1 - t}
q^{[uk^{2}+(v+2ut)k]/2}, \qquad t \in \mathbb{Z}, \label{la somme
sur k}
          \end{gather}
with $q$ to be formally replaced by $ {e} ^{2 \pi {i}/{w}}$.
Moreover, as a more general version of (\ref{96man}), we have
          \[
uw^{2}+vw = (uw+v)w \equiv 0 \  \ ({\rm mod} \ 2w),
          \]
which shows that, in spite of the factor 1/2 in the exponent of
$q$, a translation by $w$ of any of the indices $k$ does not
modify the sum in (\ref{la somme sur k}). Hence, (\ref{la somme
sur k}) leads to
          \begin{gather}
   S(u,v,w) = q^{(ut^{2}+vt)/2} S(u,v+2ut,w).
   \label{General relation}
          \end{gather}
For $t$ ranging and f\/ixed $u$, $v$, and $w$, the number of
dif\/ferent values of $v+2ut$ modulo $2(2j+1)$ is $\vert w \vert /
\gcd (u,w)$; the corresponding Gauss sums are equal up to a phase
factor. We now give two applications of formula (\ref{General
relation}).

First, for $u$ and $n$ integers and $w$ odd integer, we obtain
          \begin{gather}
S(u,2n-uw,w)=q^{-(w-1)(w+1)u / 8 + (w-1) n / 2} S(u,2n-u,w)
   \label{First derived relation}
          \end{gather}
as a particular case of (\ref{General relation}). In fact, one can
show that for $w$ odd, (\ref{First derived relation}) is
equivalent to the general formula~(\ref{General relation}).

As a second application of (\ref{General relation}), note that if
there exists $t$ such that $ut+v \equiv 0$ modulo $w$, then
(\ref{General relation}) yields
          \begin{gather}
   S(u,v,w) = \pm S(u,-v,w).
   \label{SpmS}
          \end{gather}
The minus sign may occur solely when $u$, $v$, and $w$ are even.
To see when it ef\/fectively occurs, let us consider the equation
in $t$
          \begin{gather}
    ut+v \equiv w \ ({\rm mod} \ 2w)
    \label{tartampion}
          \end{gather}
and let $v_{2}$ be the 2-valuation of integers. If $v_{2}(u) \leq
v_{2}(w)$ and (\ref{tartampion}) has an odd solution, or if
$v_{2}(u) \geq v_{2}(w)+1$ and (\ref{tartampion}) has a solution,
then there is a minus sign. Otherwise there is a plus sign. For
f\/ixed $u$ and $w$, the number of values of $v$ for which
(\ref{SpmS}) appears with a minus sign is $\vert w \vert / \gcd
(u,w)$. Now, by using again the translation method, we can show
that
          \begin{gather}
   S(u,v,w) = \sum_{k=- \vert w \vert +1}^{0} q^{(uk^{2}+vk)/2}
            = S(u,-v,w),
   \label{SpS}
          \end{gather}
a result that also follows by applying twice the reciprocity
theorem \cite{22Berndt} for generalized quadratic Gauss sums
whenever $u \neq 0$. A comparison of (\ref{SpmS}) and (\ref{SpS})
shows that certain Gauss sums $S(u,v,w)$ vanish. One may check
this numerically for ($u=2$; $v = 2, 6, 10, 14$; $w=8$) and
($u=4$; $v = 2, 6, 10    $; $w=6$).

\section{On a Gaussian sum}

As a corollary of Proposition 8, the sum rule
          \begin{gather}
\left\vert \sum_{k=0}^{d-1}  {e} ^{{i} {\pi} [k (d-k) \lambda + 2
k \mu] / d} \right\vert = \sqrt{d} \label{Gaussian}
          \end{gather}
holds for $d$ prime, $\vert \lambda \vert = 1, 2, \dots, d-1$ and
                     $\vert \mu     \vert = 0, 1, \dots, d-1$. The proof of
(\ref{Gaussian}) follows from the introduction of $k=j+m$ and
$d=2j+1$ in (\ref{overlap rasb}), (\ref{definition of rho}), and
(\ref{module de l'overlap pour 2j+1 premier}). Equation
(\ref{Gaussian}) can be derived also by adapting the results of
Exercises~23 (p.~47) and~12 (p.~44) of \cite{22Berndt} (a result
kindly communicated to the authors by B.C.~Berndt and R.J.~Evans).

\subsection*{Acknowledgements}
The senior author (M.R.K.) acknowledges Philippe Langevin for
useful correspondence. The authors thank Hubert de Guise, Michel
Planat, and Metod Saniga for interesting discussions. They are
indebted to Bruce Berndt and Ron Evans for providing them with an
alternative proof of the result in Appendix C. Thanks are due to the Referees for useful
 and constructive suggestions.

\pdfbookmark[1]{References}{ref}
\LastPageEnding


\begin{thebibliography}{99}

\footnotesize\itemsep=0pt

\bibitem{1Melvin} Melvin M.A.,
Simplif\/ication in f\/inding symmetry-adapted eigenfunctions,
{\it Rev. Modern Phys.} {\bf 28} (1956), 18--44.


\bibitem{2Altmann-1}
Altmann S.L., Cracknell A.P., Lattice harmonics I. Cubic groups,
{\it Rev. Modern Phys.} {\bf 37} (1965), 19--32.


\bibitem{2Altmann-2}
Altmann S.L., Bradley C.J., Lattice harmonics II. Hexagonal
close-packed lattice, {\it Rev. Modern Phys.} {\bf 37} (1965),
33--45.


\bibitem{3Kibler-1}
Kibler M., Ionic and paramagnetic energy levels: algebra. I, {\it
J. Mol. Spectrosc.}        {\bf 26} (1968), 111--130.


\bibitem{3Kibler-2}
Kibler M., Energy levels of paramagnetic ions: algebra. II, {\it
Int. J. Quantum Chem.}     {\bf  3} (1969), 795--822.


\bibitem{4Moret-1}
Michelot F., Moret-Bailly J., Expressions alg\'ebriques
approch\'ees de symboles de couplage et de fonctions de base
adapt\'es \`a la sym\'etrie cubique, {\it J. Phys. (Paris)} {\bf
36} (1975), 451---460.


\bibitem{4Moret-2}
Champion J.P., Pierre G., Michelot F., Moret-Bailly J.,
Composantes cubiques normales des tenseurs sph\'eriques, {\it Can.
J. Phys.} {\bf 55} (1977), 512--520.


\bibitem{5Patera-1}
Patera J., Winternitz P., A new basis for the representations of
the rotation group. Lam\'e and Heun polynomials, {\it J. Math.
Phys.} {\bf 14} (1973), 1130--1139.


\bibitem{5Patera-2}
Patera J., Winternitz P., On bases for irreducible representations
of O(3) suitable for systems with an arbitrary f\/inite symmetry
group, {\it J. Chem. Phys.} {\bf 65} (1976), 2725--2731.


\bibitem{5Patera-3}
Michel L., Invariants polynomiaux des groupes de sym\'etrie
mol\'eculaire et cristallographique, in Group Theoretical Methods
in Physics, Editors R.T.~Sharp and B.~Kolman, Academic Press, New
York, 1977, 75--91.


\bibitem{6KibPla}
Kibler M.R., Planat M., A SU(2) recipe for mutually unbiased
bases, {\it Internat. J. Modern Phys.~B} {\bf 20} (2006),
1802--1807,
\href{http://arxiv.org/abs/quant-ph/0601092}{quant-ph/0601092}.


\bibitem{7quons-1}
Arik M., Coon D.D., Hilbert spaces of analytic functions and
generalized coherent states, {\it J. Math. Phys.} {\bf 17} (1976),
524--527.


\bibitem{7quons-2}
Biedenharn L.C., The quantum group SU$_q$(2) and a $q$-analogue of
the boson operators, {\it J. Phys. A: Math. Gen.} {\bf 22} (1989),
L873--L878.


\bibitem{7quons-3}
Sun C.-P., Fu H.-C., The $q$-deformed boson realisation of the
quantum group SU($n$)$_q$ and its representations, {\it J. Phys.
A: Math. Gen.} {\bf 22} (1989), L983--L986.


\bibitem{7quons-4}
Macfarlane A.J., On $q$-analogues of the quantum harmonic
oscillator and the quantum group SU(2)$_q$, {\it J.~Phys.~A: Math.
Gen.} {\bf 22} (1989), 4581--4588.


\bibitem{7quons-5}
Daoud M., Kibler M., Fractional supersymmetric quantum mechanics
as a set of replicas of ordinary supersymmetric quantum mechanics,
{\it Phys. Lett. A} {\bf 321} (2004), 147--151,
\href{http://arxiv.org/abs/math-ph/0312019}{math-ph/0312019}.


\bibitem{8DaoHasKib-1}
Daoud M., Hassouni Y., Kibler M., The $k$-fermions as objects
interpolating between fermions and bosons, in Symmetries in
Science X, Editors B.~Gruber and M.~Ramek, Plenum Press, New York,
1998, 63--67,
\href{http://arxiv.org/abs/quant-ph/9710016}{quant-ph/9710016}.


\bibitem{8DaoHasKib-2}
Daoud M., Hassouni Y., Kibler M., Generalized supercoherent
states, {\it Phys. Atom. Nuclei} {\bf 61} (1998), 1821--1824,
\href{http://arxiv.org/abs/quant-ph/9804046}{quant-ph/9804046}.


\bibitem{8DaoHasKib-3}
Daoud M., Kibler M.R., Fractional supersymmetry and hierarchy of
shape invariant potentials, {\it J. Math. Phys.} {\bf 47} (2006),
122108, 11~pages,
\href{http://arxiv.org/abs/quant-ph/0609017}{quant-ph/0609017}.


\bibitem{9KibDao-1}
Kibler M., Daoud M., Variations on a theme of quons: I. A non
standard basis for the Wigner--Racah algebra of the group SU(2),
{\it Recent Res. Devel. Quantum Chem.} {\bf  2} (2001), 91--99.


\bibitem{9KibDao-2}
Daoud M., Kibler M., Fractional supersymmetric quantum mechanics,
{\it Phys. Part. Nuclei (Suppl.~1)} {\bf 33} (2002), S43--S51.


\bibitem{10Schwinger}
Schwinger J., On angular momentum, in Quantum Theory of Angular
Momentum, Editors L.C. Biedenharn and H. van Dam, Academic Press,
New York, 1965, 229--279.


\bibitem{11JMLL}
L\'evy-Leblond J.-M., Azimuthal quantization of angular momentum,
{\it Rev. Mex. F\'\i s.} {\bf 22} (1973), 15--23.


\bibitem{12Ellinas-1}
Chaichian M., Ellinas D., On the polar decomposition of the
quantum SU(2) algebra, {\it J. Phys. A: Math. Gen.}  {\bf 23}
(1990), L291--L296.


\bibitem{12Ellinas-2}
Ellinas D., Phase operators via group contraction, {\it J. Math.
Phys.} {\bf 32} (1991), 135--141.


\bibitem{12Ellinas-3}
Ellinas D., Quantum phase angles and su($\infty$), {\it J. Mod.
Opt.} {\bf 38} (1991), 2393--2399.


\bibitem{12Ellinas-4}
Ellinas D., Floratos E.G., Prime decomposition and correlation
measure of f\/inite quantum systems, {\it J.~Phys.~A: Math. Gen.}
{\bf 32} (1999), L63--L69,
\href{http://arxiv.org/abs/quant-ph/9806007}{quant-ph/9806007}.


\bibitem{13FFZ}
Fairlie D.B., Fletcher P., Zachos C.K., Inf\/inite-dimensional
algebras and a trigonometric basis for the classical Lie algebras,
{\it J. Math. Phys.} {\bf 31} (1990), 1088--1094.


\bibitem{14PateraZassenhaus}
Patera J., Zassenhaus H., The Pauli matrices in $n$ dimensions and
f\/inest gradings of simple Lie algebras of type $A_{n-1}$, {\it
J. Math. Phys.} {\bf 29} (1988), 665--673.


\bibitem{14bisWeyl}
Weyl H., The theory of groups and quantum mechanics, Dover
Publications, New York, 1950.


\bibitem{15Delsarte}
Delsarte P., Goethals J.M., Seidel J.J., Bounds for systems of
lines and Jacobi polynomials, {\it Philips Res. Repts.} {\bf 30}
(1975), 91--105.


\bibitem{16Ivanovic}
Ivanovi\'c I.D., Geometrical description of quantum state
determination, {\it J. Phys. A: Math. Gen.} {\bf 14} (1981),
3241--3245.


\bibitem{17lesWootters-1}
Wootters W.K., Quantum mechanics without probability amplitudes,
{\it Found. Phys.} {\bf 16} (1986), 391--405.


\bibitem{17lesWootters-2}
Wootters W.K., A Wigner-function formulation of f\/inite-state
quantum mechanics, {\it Ann. Phys. (N.Y.)} {\bf 176} (1987),
1--21.


\bibitem{17lesWootters-3}
Wootters W.K., Fields B.D., Optimal state-determination by
mutually unbiased measurements, {\it Ann. Phys. (N.Y.)} {\bf 191}
(1989), 363--381.


\bibitem{17lesWootters-4}
Gibbons K.S., Hof\/fman M.J., Wootters W.K., Discrete phase space
based on f\/inite f\/ields, {\it Phys. Rev. A} {\bf 70} (2004),
062101, 23~pages,
\href{http://arxiv.org/abs/quant-ph/0401155}{quant-ph/0401155}.


\bibitem{17lesWootters-5}
Wootters W.K., Picturing qubits in phase space, {\it IBM J. Res.
Dev.} {\bf 48} (2004), 99, 12~pages,
\mbox{\href{http://arxiv.org/abs/quant-ph/0306135}{quant-ph/0306135}}.


\bibitem{17lesWootters-6}
Wootters W.K., Quantum measurements and f\/inite geometry, {\it
Found. Phys.}  {\bf 36} (2006), 112--126,
\mbox{\href{http://arxiv.org/abs/quant-ph/0406032}{quant-ph/0406032}}.


\bibitem{18Calderbank}
Calderbank A.R., Cameron P.J., Kantor W.M., Seidel J.J., Zopf
4-Kerdock codes orthogonal spreads and extremal euclidean
line-sets, {\it Proc. London Math. Soc.} {\bf 75} (1997),
436--480.


\bibitem{19Edmonds}
Edmonds A.R., Angular momentum in quantum mechanics, Princeton
University Press, Princeton, 1960.


\bibitem{20Racah}
Racah G., Theory of complex spectra. II, {\it Phys. Rev.} {\bf 62}
(1942), 438--462.


\bibitem{21GreKib-1}
Grenet G., Kibler M., On the operator equivalents, {\it Phys.
Lett. A}  {\bf 68} (1978), 147--150.


\bibitem{21GreKib-2}
Kibler M., Grenet G., On the SU$_2$ unit tensor, {\it J. Math.
Phys.} {\bf 21} (1980), 422--439.


\bibitem{22Berndt}
Berndt B.C., Evans R.J., Williams K.S., Gauss and Jacobi sums,
Wiley, New York, 1998.


\bibitem{23Bandyo}
Bandyopadhyay S., Boykin P.O., Roychowdhury V., Vatan F., A new
proof for the existence of mutually unbiased bases, {\it
Algorithmica} {\bf 34} (2002), 512--528,
\href{http://arxiv.org/abs/quant-ph/0103162}{quant-ph/0103162}.


\bibitem{24Lawrence}
Lawrence J., Brukner \v{C}., Zeilinger A., Mutually unbiased
binary observable sets on $N$ qubits, {\it Phys. Rev.~A} {\bf 65}
(2002), 032320, 5~pages,
\href{http://arxiv.org/abs/quant-ph/0104012}{quant-ph/0104012}.


\bibitem{25Klimov-1}
Klimov A.B., S\'anchez-Soto L.L., de Guise H., Multicomplementary
operators via f\/inite Fourier transform, {\it J. Phys. A: Math.
Gen.} {\bf 38} (2005), 2747--2760,
\href{http://arxiv.org/abs/quant-ph/0410155}{quant-ph/0410155}.


\bibitem{25Klimov-2}
Klimov A.B., S\'anchez-Soto L.L., de Guise H., A
complementary-based approach to phase in f\/inite-dimensional
quantum systems, {\it J. Opt. B: Quantum Semiclass. Opt.} {\bf 7}
(2005), 283--287,
\href{http://arxiv.org/abs/quant-ph/0410135}{quant-ph/0410135}.


\bibitem{25Klimov-3}
Romero J.L., Bj\"ork G., Klimov A.B., S\'anchez-Soto L.L.,
Structure of the sets of mutually unbiased bases for $N$ qubits,
{\it Phys. Rev. A} {\bf 72} (2005), 062310, 8~pages,
\href{http://arxiv.org/abs/quant-ph/0508129}{quant-ph/0508129}.


\bibitem{26LeBellac}
Le Bellac M., Physique quantique, EDP Sciences/CNRS \'Editions,
Paris, 2003.


\bibitem{27Chaturvedi}
Chaturvedi S., Aspects of mutually unbiased bases in
odd-prime-power dimensions, {\it Phys. Rev. A} {\bf 65} (2002),
044301, 3~pages,
\href{http://arxiv.org/abs/quant-ph/0109003}{quant-ph/0109003}.


\bibitem{28Pittinger-1}
Pittenger A.O., Rubin M.H., Mutually unbiased bases, generalized
spin matrices and separability, {\it Linear Algebr. Appl.} {\bf
390} (2004), 255--278,
\href{http://arxiv.org/abs/quant-ph/0308142}{quant-ph/0308142}.


\bibitem{28Pittinger-2}
Pittenger A.O., Rubin M.H., Wigner functions and separability for
f\/inite systems, {\it J. Phys. A: Math. Gen.} {\bf 38} (2005),
6005--6036,
\href{http://arxiv.org/abs/quant-ph/0501104}{quant-ph/0501104}.


\bibitem{29SPR-1}
Saniga M., Planat M., Rosu H., Mutually unbiased bases and
f\/inite projective planes, {\it J. Opt. B: Quantum Semiclass.
Opt.} {\bf 6} (2004), L19--L20,
\href{http://arxiv.org/abs/math-ph/0403057}{math-ph/0403057}.


\bibitem{29SPR-2}
Saniga M., Planat M., Sets of mutually unbiased bases as arcs in
f\/inite projectives planes, {\it Chaos Solitons Fractals} {\bf
26} (2005), 1267--1270,
\href{http://arxiv.org/abs/quant-ph/0409184}{quant-ph/0409184}.


\bibitem{29SPR-2bis}
Saniga M., Planat M., Hjelmslev geometry of mutually unbiased
bases, {\it J. Phys. A: Math. Gen.} {\bf 39} (2006), 435--440,
\href{http://arxiv.org/abs/math-ph/0506057}{math-ph/0506057}.


\bibitem{29SPR-3}
Planat M., Rosu H., Mutually unbiased phase states, phase
uncertainties, and Gauss sums, {\it Eur. Phys.~J.~D} {\bf 36}
(2005), 133--139.


\bibitem{29SPR-4}
Planat M., Saniga M., Galois algebras of squeezed quantum phase
states, {\it J. Opt. B: Quantum Semiclass. Opt.} {\bf 7} (2005),
S484--S489.


\bibitem{29SPR-5}
Saniga M., Planat M., Finite geometries in quantum theory: from
Galois (f\/ields) to Hjelmslev (rings), {\it Internat. J. Modern
Phys.~B} {\bf 20} (2006), 1885--1892.


\bibitem{30Vourdas-1}
Vourdas A., SU(2) and SU(1,1) phase states, {\it Phys. Rev.~A}
{\bf 41} (1990), 1653--1661.


\bibitem{30Vourdas-2}
Vourdas A., The angle-angular momentum quantum phase space, {\it
J. Phys. A: Math. Gen.} {\bf 29} (1996), 4275--4288.


\bibitem{30Vourdas-3}
Vourdas A., Quantum systems with f\/inite Hilbert space, {\it Rep.
Prog. Phys.} {\bf 67} (2004), 267--320.


\bibitem{30Vourdas-4}
Vourdas A., Galois quantum systems, {\it J. Phys. A: Math. Gen.}
{\bf 38} (2005), 8453--8471.


\bibitem{30Vourdas-5}
Vourdas A., Galois quantum systems, irreducible polynomials and
Riemann surfaces, {\it J. Math. Phys.}   {\bf 47} (2006), 092104,
15~pages.


\bibitem{31Klappenecker-1}
Klappenecker A., R\"otteler M., Constructions of mutually unbiased
bases, {\it Lect. Notes Comput. Sci.} {\bf 2948} (2004), 137--144,
\href{http://arxiv.org/abs/quant-ph/0309120}{quant-ph/0309120}.


\bibitem{31Klappenecker-2}
Klappenecker A., R\"otteler M., Shparlinski I.E., Winterhof A., On
approximately symmetric informationally complete positive
operator-valued measures and related systems of quantum states,
{\it J. Math. Phys.} {\bf 46} (2005), 082104, 17~pages,
\href{http://arxiv.org/abs/quant-ph/0503239}{quant-ph/0503239}.


\bibitem{32Grassl}
Grassl M., On SIC-POVMs and MUBs in dimension 6, in Proceedings
ERATO Conference on Quantum Information Science (EQIS 2004),
Editor J. Gruska, Tokyo, 2005, 60--61,
\href{http://arxiv.org/abs/quant-ph/0406175}{quant-ph/0406175}.


\bibitem{33Durt-1}
Durt T., If 1=2+3, then 1=2.3: Bell states, f\/inite groups, and
mutually unbiased bases, a unifying approach, {\it J. Phys. A:
Math. Gen.} {\bf 38} (2005), 5267--5283,
\href{http://arxiv.org/abs/quant-ph/0401046}{quant-ph/0401046}.


\bibitem{33Durt-2}
Colin S., Corbett J., Durt T., Gross D., About SIC POVMs and
discrete Wigner distributions, {\it J. Opt. B: Quantum Semiclass.
Opt.} {\bf 7} (2005), S778--S785.


\bibitem{33Durt-3}
Durt T., About the mean king's problem and discrete Wigner
distributions, {\it Internat. J. Modern Phys.~B} {\bf 20} (2006),
1742--1760.


\bibitem{34Wocjan}
Wocjan P., Beth T., New construction of mutually unbiased bases in
square dimensions, {\it Quantum Inf. Comput.} {\bf 5} (2005),
93--101,
\href{http://arxiv.org/abs/quant-ph/0407081}{quant-ph/0407081}.


\bibitem{35Archer}
Archer C., There is no generalization of known formulas for
mutually unbiased bases, {\it J. Math. Phys.} {\bf 46} (2005),
022106, 11~pages,
\href{http://arxiv.org/abs/quant-ph/0312204}{quant-ph/0312204}.


\bibitem{36Bengtsson-1}
Bengtsson I., MUBs, polytopes, and f\/inite geometries,
\href{http://arxiv.org/abs/quant-ph/0406174}{quant-ph/0406174}.


\bibitem{36Bengtsson-2}
Bengtsson I., Ericsson \AA., Mutually unbiased bases and the
complementary polytope, {\it Open Syst. Inf. Dyn.} {\bf 12}
(2005), 107--120,
\href{http://arxiv.org/abs/quant-ph/0410120}{quant-ph/0410120}.


\bibitem{36Bengtsson-3}
Bengtsson I., Bruzda W., Ericsson \AA.,
                       Larsson J.-\AA., Tadej W., \.{Z}yckowski K.,
               Mubs and Hadamards of order six,
               \href{http://arxiv.org/abs/quant-ph/0610161}{quant-ph/0610161}.


\bibitem{36Bengtsson-4}
Bengtsson I., Three ways to look at mutually unbiased bases,
\href{http://arxiv.org/abs/quant-ph/0610216}{quant-ph/0610216}.


\bibitem{37Boykin}
Boykin P.O., Sitharam M., Tiep P.H., Wocjan P., Mutually unbiased
bases and orthogonal decompositions of Lie algebras,
\href{http://arxiv.org/abs/quant-ph/0506089}{quant-ph/0506089}.


\bibitem{38Hayashi}
Hayashi A., Horibe M., Hashimoto T., Mean king's problem with
mutually unbiased bases and orthogonal Latin squares, {\it Phys.
Rev. A} {\bf 71} (2005), 052331, 4~pages,
\href{http://arxiv.org/abs/quant-ph/0502092}{quant-ph/0502092}.

\end{thebibliography}
\end{document}